\documentclass[12pt,epsf]{article}
\usepackage{amssymb,amsmath,amsbsy,amsthm}
\usepackage[dvipsnames]{xcolor}
\usepackage[utf8]{inputenc}
\usepackage{footmisc}
\usepackage{graphicx,color}

\newcommand{\ben}{\begin{enumerate}}
\newcommand{\een}{\end{enumerate}}

\newcommand{\be}{\begin{equation}}
\newcommand{\ee}{\end{equation}}
\newcommand{\bea}{\begin{eqnarray}}
\newcommand{\eea}{\end{eqnarray}}
\newcommand{\lb}{\left(}
\newcommand{\rb}{\right)}
\newcommand{\la}{\langle}
\DeclareMathOperator{\sign}{sign}

\newcommand{\ra}{\rangle}
\newcommand{\nn}{\nonumber}
\newcommand{\lam}{\lambda}
\newcommand{\p}{\partial}
\newcommand{\Om}{\Omega}
\newcommand{\mO}{{\mathcal O}}

\newcommand{\nbox}{{\,\lower0.9pt\vbox{\hrule \hbox{\vrule height 0.2 cm \hskip 0.19 cm \vrule height 0.2 cm}\hrule}\,}}

\def\href#1#2{#2}

\begin{document}
\begin{titlepage}
\vbox{
    \halign{#\hfil         \cr
           } 
      }  
\vspace*{15mm}
\begin{center}

{\large \bf Constraining Quantum Fields using Modular Theory}

\vspace*{15mm}
\vspace*{1mm}
Nima Lashkari
\vspace*{1cm}
\let\thefootnote\relax\footnote{$\mathrm{lashkari@ias.edu}$}

{\it{ School of Natural Sciences, Institute for Advanced Study,\\
Einstein Drive, Princeton, NJ, 08540, USA}\\
\vspace*{0.1cm}}

\vspace*{0.7cm}
\end{center}
\begin{abstract}
Tomita-Takesaki modular theory provides a set of algebraic tools in quantum field theory that is suitable for the study of the information-theoretic properties of states. For every open set in spacetime and choice of two states, the modular theory defines a positive operator known as the relative modular operator that decreases monotonically under restriction to subregions. We study the consequences of this operator monotonicity inequality for correlation functions in quantum field theory. We do so by constructing a one-parameter R{\'e}nyi family of information-theoretic measures from the relative modular operator that inherit monotonicity by construction and reduce to correlation functions in special cases. In the case of finite quantum systems, this R{\'e}nyi family is the sandwiched R{\'e}nyi divergence and we obtain a new simple proof of its monotonicity.
Its monotonicity implies a class of constraints on correlation functions in quantum field theory, only a small set of which were known to us. We explore these inequalities for free fields and conformal field theory. We conjecture that the second null derivative of R{\'e}nyi divergence is non-negative which is a generalization of the quantum null energy condition to the R{\'e}nyi family.
\end{abstract}

\end{titlepage}

\vskip 1cm

\section{Introduction}\label{sec:intro}

In recent years, the tools and techniques of quantum information theory have found many applications in many-body local quantum systems from condensed matter to field theory and gravity. In relativistic quantum field theory, studying the constraints of unitarity and causality on information-theoretic measures has led to the discovery of new inequalities that hold universally in all theories. The strong sub-additivity of entanglement entropy has been used to derive c-theorems in various dimensions \cite{casini2004finite,casini2012renormalization,casini2017markov}. Strong sub-additivity is a special case of the monotonicity of relative entropy which has been used to prove the average null energy condition (ANEC) in flat space and the generalized second law in curved spacetimes \cite{faulkner2016modular,wall2012proof}. In this work, we explore a large class of monotonicity constraints of information-theoretic measures in quantum field theory.

In quantum field theory, there are no density matrices associated to finite regions of spacetime. 
This is an obstacle on the way of applying the entanglement theory to quantum fields which reflects itself as ultraviolet divergences that appear in entanglement entropy. In some cases, one can use scaling arguments to isolate universal pieces in the calculations. These ultraviolet divergences point to the fact that entanglement in quantum field theory is a property of the algebra and not the states \cite{Witten:2018zxz}. 

A more satisfactory approach is provided by the Tomita-Takesaki modular theory that studied the algebra of local operators in quantum field theory; see \cite{Haag:1992hx, Witten:2018zxz} for a review. Modular theory connects with information theory by defining a positive operator known as the relative modular operator. Araki realized that the expectation value of the logarithm of this operator generalizes the density matrix definition of relative entropy to arbitrary quantum systems including quantum field theory \cite{araki1976relative}. Relative entropy is a distance measure central to quantum information theory from which most other entanglement measures can be derived; see \cite{vedral2002role} for a review.

In modular theory, the relative modular operator $\Delta^A_{\Omega|\Phi}$ is defined for every spacetime region $A$ and choice of two states $|\Omega\ra$ and $|\Phi\ra$. The locality of quantum field theory imposes a strong constraint on how the relative modular operator changes as one changes the region $A$. Most importantly for us, the relative modular operator is monotonic, that is to say for any region $A\subset B$ we have
\bea
\Delta^A_{\Omega|\Phi}\geq \Delta^B_{\Omega|\Phi}
\eea
as an operator inequality. It implies the monotonicity of relative entropy, however, as an operator inequality, it is much stronger. In this work, we study the consequences of this inequality in quantum field theory.

In section \ref{sec:modular}, we review the modular theory emphasizing those properties of the relative modular operator that play an important role in our work. In section \ref{sec:petz}, we introduce the Petz divergence as an information-theoretic measure that monotonically increases as one enlarges the region, and show that it satisfies all the desirable information-theoretic properties one expects from a R{\'e}nyi generalization of relative entropy.
 In section \ref{sec:sand}, motivated by the work of Araki and Masuda in \cite{araki1982positive} we define an alternative R{\'e}nyi family for relative entropy and show that it is the generalization of the sandwiched R{\'e}nyi divergence to quantum field theory. Similar discussions appear in \cite{jaksic2011entropic,jencova2016r,berta2018renyi}. Our approach has the advantage that we do not need to resort to the proofs in \cite{frank2013monotonicity,beigi2013sandwiched,berta2018renyi} to show the monotonicity of sandwiched R{\'e}nyi divergences. It is implied by the monotonicity of relative modular operator. For completeness, we review of the proof of the monotonicity of relative modular operator with an emphasis on finite dimensional Hilbert spaces in section \ref{sec:finite}.

In \cite{lashkari2014relative} a Euclidean path-integral was constructed that computes sandwiched R{\'e}nyi divergences, and it was shown that for certain values of the R{\'e}nyi index they are given by correlation functions. 
Section \ref{sec:constraint} starts with this path-integral construction and explores the consequences of monotonicity for these correlation functions. We find new inequalities that the $2n$-point correlation functions have to satisfy as a consequence of monotonicity. They can be interpreted as the non-negativity of certain off-diagonal elements of null momentum. See the inequality in (\ref{constraint}) for a concrete example.

Studying the dependence of these $2n$-point correlation functions on the region size we are naturally led to the conjecture that the second null (or spacelike) derivatives of the sandwiched R{\'e}nyi divergence is non-negative. This can be thought of as a generalization of quantum null energy condition to the R{\'e}nyi family. We provide evidence for the conjecture using reflection positivity of Euclidean correlators. We perform sample calculations of the sandwiched R{\'e}nyi divergence and its derivatives in shape deformations in free two-dimensional bosons and higher-dimensional conformal field theories in a local quench limit.

Finally, we conclude in section \ref{sec:Dis} by discussing generalizations and comment on potential implications of our work for constraining bulk effective field theories in holography.

\section{Tomita-Takesaki modular theory}\label{sec:modular}

A general quantum system is described by an algebra of bounded operators with a $\dagger$ operation and a notion of a norm. In this abstract language, states are linear maps from the algebra to correlation functions. Given a state, there is a well-defined prescription to construct an irreducible representation of such an algebra in the Hilbert space.\footnote{This prescription is called the Gelfand–Naimark–Segal construction. See \cite{Haag:1992hx} for a review.} If the algebra is generated by operators that satisfy the canonical commutation relation Lie algebra then such an irreducible representation is unique up to unitary equivalence.\footnote{This is called the Stone-von Neumann theorem.} This is not the case in quantum field theory. There are many inequivalent representations of the algebra of operators in the Hilbert space that correspond to different superselection sectors. This is one of the main motivations to study the local algebra of operators in quantum field theory.\footnote{By the local algebra we mean the algebra associated with a region of spacetime.}

The algebra of local operators in quantum field theory differs from finite dimensional quantum systems in essential ways. Given a state living on a Cauchy slice, there is no notion of density matrix operator corresponding to a subregion. This causes a problem for a naive generalization of entanglement theory from finite quantum systems to field theory. The Tomita-Takesaki modular theory allows us to define information-theoretic measures in quantum field theory without the need to define density matrices.

Consider a relativistic quantum field theory in Minkowski space in $D$ spacetime dimensions. For every open set $A$ in spacetime, there exists a von Neumann algebra of bounded operators $\mathcal{A}_A$ that is generated by bounded functions of fundamental fields integrated against test functions supported on $A$. Locality requires that $\mathcal{A}_B\subset \mathcal{A}_A$ if $B\subset A$.\footnote{This is often called the Isotony condition.}   
The closure of the union of algebras associated with all open sets forms the global algebra of quantum field theory.
In a relativistic quantum field theory, a Poincar{\'e} transformation that sends $A\to \Lambda(A)$ corresponds to an automorphism of the global algebra that transforms $\mathcal{A}\to \mathcal{A}_{\Lambda(A)}$. We think of states on $A$ as linear maps $\omega_A$ from the local algebra $\mathcal{A}_A$ to expectation values, and denote the states of the global algebra with vectors $|\Om\ra$ in the Hilbert space. 

Every von Neumann algebra $\mathcal{A}_A$ has a notion of $\dagger$. Therefore, for any two vectors $|\Om\ra$ and $|\Phi\ra$ and a region $A$ one can define the relative Tomita operator by its action
\bea\label{relTomita}
&&S^A_{\Phi|\Om} a|\Om\ra=a^\dagger|\Phi\ra,
\eea 
where $a$ is an operators in the algebra $\mathcal{A}_A$  of region $A$, and $|\Omega\ra$ is carefully chosen such that $a|\Om\ra$ and $a'|\Omega\ra$ are dense in the Hilbert space.\footnote{The relative Tomita operator is the closure of the operator defined in (\ref{relTomita}). In this work, we use the notation $S_{\Phi|\Om}$ to mean the closed operator.}  We further assume that there exists no $a\in\mathcal{A}$ such that $a|\Om\ra=0$. Such states are called cyclic and separating. The definition in (\ref{relTomita}) can be extended to arbitrary vectors, however, to simplify the presentation we restrict our discussion to vectors that are cyclic and separating. 

The relative modular operator $\Delta_{\Phi|\Omega}$ is the norm of the relative Tomita operator:
\bea
\Delta_{\Phi|\Om}\equiv S^\dagger_{\Phi|\Om}S_{\Phi|\Om}\ .
\eea
It is unbounded and manifestly non-negative. From the definition of the Tomita operator it is clear that 
\bea
&&S_{\Phi|\Om}S_{\Om|\Phi}=1, \qquad S_{\Phi|\Om}^\dagger S_{\Om|\Phi}^\dagger=1,
\eea
hence the action of $\Delta^{-1}_{\Om|\Phi}=S_{\Phi|\Om}S_{\Phi|\Om}^\dagger$. Furthermore, the Tomita operator for region $A$ is the Hermitian conjugate of that of $A'$ (its causal complement): $(S^A_{\Phi|\Om})^\dagger=S^{A'}_{\Phi|\Om}$ \cite{jones}. As a result, the relative modular operators $(\Delta^A_{\Om|\Phi})^{-1}=\Delta^{A'}_{\Phi|\Om}$.\footnote{We have assumed that the algebra of $A$ and $A'$ commute and have no operators in common except for the identity operator. Such a von Neumann algebra is called a ``factor".}  This operator plays an important role in our discussion.
Most importantly for us, the relative modular operator for any two vectors $|\Phi\ra$ and $|\Om\ra$ satisfies the monotonicity inequality
\bea\label{monotone}
\Delta^B_{\Om|\Phi}\leq \Delta^A_{\Om|\Phi}
\eea 
for any $\mathcal{A}_A\subset \mathcal{A}_B$. This is a consequence of the fact that the domain of $S^B_{\Phi|\Om}$ is an extension of that of $S^A_{\Phi|\Om}$.\footnote{See \cite{Witten:2018zxz} for a proof based on the ideas presented in \cite{borchers2000revolutionizing}. For completeness, in section \ref{sec:finite} we review a variant of this proof adapted for finite-dimensional systems presented in \cite{nielsen2004simple}.} That is to say for any $a\in\mathcal{A}_A$ we have
\bea
&&S^A_{\Om|\Phi}a|\Phi\ra=S^B_{\Om|\Phi}a|\Phi\ra=a^\dagger|\Om\rangle\ .\eea
This implies that for vectors $a|\Phi\ra$ with $a\in\mathcal{A}_A$ the monotonicity of the relative modular operator is saturated:
\bea
&&\la \Phi|a^\dagger \Delta^A_{\Om|\Phi}a|\Phi\ra=\la \Phi|a^\dagger \Delta^B_{\Om|\Phi}a|\Phi\ra\ .
\eea
To obtain an inequality we consider non-linear functions of $\Delta_{\Om|\Phi}$ that inherit monotonicity by construction. A function $f:(0,\infty)\to (0,\infty)$ is called operator monotone if for any self-adjoint operators $X\geq Y\geq 0$ we have $f(X)\geq f(Y)\geq 0$. An important example is $f(z)=z^\alpha$ which is operator monotone for 
for $0<\alpha<1$ \cite{Witten:2018zxz}. For invertible operators, $X>Y>0$ implies $Y^{-1}>X^{-1}$.\footnote{The inequality $X>Y>0$ implies $1>X^{-1/2}Y X^{-1/2}$. The operator $X^{-1/2}Y X^{-1/2}$ is non-negative and has a  spectral decomposition, therefore $X^{1/2}Y^{-1}X^{1/2}>1$ and as a result we have $Y^{-1}>X^{-1}$.} The relative modular operator is invertible, therefore it satisfies the monotonicity operator inequality 
\bea\label{alphamono}
\sign(\alpha)(\Delta^A_{\Om|\Phi})^{-\alpha}\leq \sign(\alpha)(\Delta^B_{\Om|\Phi})^{-\alpha}
\eea
for $1\geq \alpha\geq -1$. 
Another important operator monotone function is the principal branch logarithm on positive definite operators which implies that Araki's relative entropy 
\bea\label{relativeentropy}
S(\Phi\|\Om)=-\la\Phi|\log\Delta_{\Om|\Phi}|\Phi\ra
\eea
is monotonic. In this work, we mostly focus on $f(z)=z^\alpha$ and briefly comment on other operator monotone functions until section \ref{sec:Dis}.


To study the entanglement between region $A$ and $A'$ it is natural to focus on quantities that are independent of unitaries in the region $A$ and $A'$. Hence, we should understand how the relative modular operator transforms when the vectors are rotated by unitaries in $A$ and $A'$. 
One can check directly from the definition in (\ref{relTomita}) that $S_{UU'\Om|VV'\Psi}=VU' S_{\Om|\Psi}V'^\dagger U^\dagger$ satisfies
\bea
S_{UU'\Om|VV'\Psi} a V V'|\Psi\ra= a^\dagger U U'|\Om\ra
\eea
for unitaries $U,V\in\mathcal{A}_A$ and $U',V'\in\mathcal{A}_{A'}$.
As a result, the relative modular operator transforms according to
\bea\label{transformunit}
\Delta_{UU'\Om|VV'\Psi}=U V'\Delta_{\Om|\Psi} V'^\dagger U^\dagger\ .
\eea
This transformation rule plays an important in section \ref{sec:petz} when we engineer information-theoretic quantities from the relative modular operator that are invariant under local unitaries.
 
In a quantum field theory with a stress tensor, shape deformations are generated by unitaries that are exponentiated integrals of the stress tensor. 
Consider a diffeomorphism that deforms region $A$ continuously to its subregion $B$ and represent it by unitary $U$. 
In this work, we consider only the shape deformations that leave the vacuum invariant:
\bea
\mathcal{A}_A=U\mathcal{A}_B U^\dagger,\qquad U|\Om\ra=|\Om\ra\ .
\eea
Examples of such unitaries are Poincare transformations and arbitrary deformations on null hypersurfaces. 
The reason is that for such unitaries, we can explicitly relate the relative modular operator of $B$ with that of $A$ using the definition of the relative Tomita operator:
\bea
&&S^{B}_{\Psi|\Om} a_B|\Om\ra= (U^\dagger S^A_{U\Psi|\Om} U) a_B |\Omega\ra
\eea
and
\bea\label{symdiff}
\Delta^{B}_{\Psi|\Om} =U^\dagger \Delta^{A}_{U\Psi|\Om} U\ .
\eea
In the limit the size of region $A$ shrinks to zero the state $|\Phi\ra$ and $|\Omega\ra$ are indistinguishable, therefore one expects the matrix elements of $f(\Delta_{\Omega|\Phi})$ to converge to those of $f(\Delta_\Phi)$. In particular, this implies that for any operator monotone function $f$ and any region $A$ we have \footnote{We thank Edward Witten for pointing this out to us.}
%
 \bea\label{fmono}
\la\Phi|f(\Delta^A_{\Om|\Phi})|\Phi\ra-f(1)\leq 0\ .
\eea
We are now ready to construct information-theoretic measures out of the relative modular operator. The class of measures that we will show to be related to the quantum field theory correlation functions is called the sandwiched R{\'e}nyi divergence. In the next section, we start by a closely related but simpler class of measures called Petz divergences.
\section{Petz divergences}\label{sec:petz}

The monotonicity of relative entropy in (\ref{alphamono}) is an operator equation which implies
\bea\label{mono}
\sign(\alpha)\la\Psi|(\Delta^A_{\Om|\Phi})^{-\alpha}|\Psi\ra\leq \sign(\alpha)\la\Psi|(\Delta^B_{\Om|\Phi})^{-\alpha}|\Psi\ra
\eea
for any $A\subset B$, $\alpha\in [-1,1]$ and any vector $|\Psi\ra$ in the Hilbert space.  The matrix element $\la\Psi|(\Delta^A_{\Om|\Phi})^{-\alpha}|\Psi\ra$ depends 
 on the information in both $A$ and the complementary region $A'$. In entanglement theory we are interested in ``semilocal" quantities, meaning those that are independent of the information content in $A'$. If $|\Phi\ra$ is cyclic and separating the vector $a|\Phi\ra$ with $a\in\mathcal{A}$ is dense in the Hilbert space, therefore it is natural to consider Petz quasi-entropies introduced in \cite{petz1985quasi}\footnote{In \cite{petz1985quasi} Petz defines the quantity in (\ref{PetzQuasi}) for a general operator monotone function $f$.}
\bea\label{PetzQuasi}
D^A_{\alpha,a}(\Phi\|\Om)=\frac{1}{\alpha}\log\la\Phi| a^\dagger (\Delta^A_{\Om|\Phi})^{-\alpha} a|\Phi\ra\ .
\eea
where the operator $a$ is normalized such that $\la\Om|a^\dagger a|\Om\ra=1$ and $\alpha\in [-1,1]$ . This normalization makes sure quasi-entropies vanish at $\alpha=-1$ because
\bea\label{quasilist}
\la\Phi| a^\dagger (\Delta^A_{\Om|\Phi}) a|\Phi\ra=\la\Om|a^\dagger a|\Om\ra=1\ .
\eea

Under unitary rotations in $A'$ the relative modular operator transforms according to (\ref{transformunit}), and since $[a,U']=0$ the Petz quasi-entropy remains invariant.

In addition, this R{\'e}nyi family has the following properties:
\begin{enumerate}

\item If $A\subset B$ and $a\in\mathcal{A}_A$ then it increases monotonically with system size: $$D^A_{\alpha,a}(\Phi\|\Om)\leq D^B_{\alpha,a}(\Phi\|\Om).$$


\item If $U$ is a unitary in $\mathcal{A}_A$ then $D_{\alpha,a}^A(U\Phi\|V\Om)=D_{\alpha,V^\dagger a U}^A(\Phi\|\Om)$. 

\item It increases monotonically in $\alpha$. If $\alpha\leq \beta$ then $D^A_{\alpha,a}(\Phi\|\Om)\leq D^A_{\beta,a}(\Phi\|\Om)$. 

\end{enumerate}

See appendix \ref{appA0} for a proof. In the remainder of this work, we focus on quasi-entropies in the case $a=1$, 
where the subregion monotonicity property extends to all subregions. For $a=1$, we have the Petz divergence:
\bea\label{Petzdiv}
&&D_\alpha(\Phi\|\Om)=\frac{1}{\alpha}\log\la\Phi|\Delta^{-\alpha}_{\Om|\Phi}|\Phi\ra
\eea
with the following additional properties:

\begin{enumerate}

\item It is non-negative and vanishes when $A$ shrinks to zero. 

\item It is invariant under the rotation of both vectors by the same unitary in $A$.

\item At $\alpha=0$ it is smooth and equal to the relative entropy.

\item Under swapping vectors $|\Phi\ra$ and $|\Om\ra$ it satisfies
\bea
D_{-\alpha}(\Phi\|\Om)=\frac{1-\alpha}{\alpha}\: D_{\alpha-1}(\Om\|\Phi)\ .
\eea
\end{enumerate}
See appendix \ref{appA} for proofs.

Petz divergences vanish at $\alpha=-1$ and increase monotonically in $\alpha$. They acquire their maximum at $\alpha=1$:
\bea
D_{1}(\Phi\|\Om)=\log\la \Phi| \Delta_{\Om|\Phi}^{-1}|\Phi\ra=\log \la \Om| S^\dagger_{\Phi|\Om}(S_{\Phi|\Om}S_{\Phi|\Om}^\dagger)S_{\Phi|\Om}|\Om\ra=\log\la\Om|\Delta^2_{\Phi|\Om}|\Om\ra\ .
\eea
If $|\Phi\ra\sim\Phi'|\Om\ra$ with $\Phi'\in\mathcal{A}'$ since $(S^A_{\Phi|\Om})^\dagger$ is an extension of $S^{A'}_{\Phi|\Om}$ we have
\bea
D_1(\Phi\|\Om)=\log\lb\frac{\la\Om||\Phi'|^4|\Om\ra}{\la \Om||\Phi'|^2|\Om\ra^2}\rb\ .
\eea

\section{Sandwiched R{\'e}nyi divergences}\label{sec:sand}

The Petz divergence satisfies all the desired properties an information-theorist would want, however it cannot be written in terms of the correlation functions of quantum field theory in a simple way. A closely related monotonic R{\'e}nyi family called the sandwiched R{\'e}nyi divergence was introduced in \cite{muller2013quantum,wilde2014strong} for finite quantum systems. In \cite{lashkari2014relative} it was shown that for special values of $\alpha$, in quantum field theory, the sandwiched R{\'e}nyi divergences can be written in terms of correlation functions. We focus on this new family in this section.

To motivate an algebraic definition of the sandwiched R{\'e}nyi divergence we follow the approach in \cite{araki1982positive}.\footnote{
A similar discussion appears in a recent work \cite{berta2018renyi}.}
Consider a vector $|\Phi\ra$ that is in the intersection of the domain of $\Delta_{\Om|\Psi}^{-1}$ for all $|\Psi\ra\in\mathcal{H}$. 
We define the sandwiched R{\'e}nyi divergence to be 
\bea\label{Renyirel1}
&&S^A_{\alpha}(\Phi\|\Om)\equiv \frac{1}{\alpha} \sup_{|\Psi\ra\in\mathcal{H}}\log\la\Phi|(\Delta_{\Om|\Psi}^A)^{-\alpha}|\Phi\ra\nn\\
&&S^A_{-\alpha}(\Phi\|\Om)\equiv \frac{-1}{\alpha} \inf_{|\Psi\ra\in\mathcal{H}}\log\la\Phi|(\Delta_{\Om|\Psi}^A)^{\alpha}|\Phi\ra
\eea
for $0\leq \alpha<1$.\footnote{We assume all vectors are normalized $\la\Psi|\Psi\ra=1$ unless mentioned otherwise.} 
If the intersection of the domains of $\Delta_{\Omega|\Psi}^{-1}$ for all $|\Psi\ra$ does not include $|\Phi\ra$ the Sandwiched R{\'e}nyi divergence is defined to be infinite. However, we will show that it is finite for a dense set of states $\Phi'|\Omega\ra$ where $\Phi'$ is an operator in the commutant.
Sandwiched R{\'e}nyi divergences satisfy all the desired properties that led us to Petz divergences. The subregion monotonicity and non-negativity are inherited from (\ref{mono}) by definition. The monotonicity in $\alpha$ follows from Holder's inequality similar to the argument presented in appendix \ref{appA}. It is also invariant under rotation by untiaries in $A'$, and
under unitary rotations in $A$, it transforms according to
$S_{\alpha}(V\Phi\|U\Omega)=S_{\alpha}(U^\dagger V\Phi\|\Om)=S_\alpha(\Phi\|V^\dagger U\Om)$. Similar to the Petz entropy it vanishes when the two vectors are the same, but as opposed to the Petz entropy, it does not vanish at $\alpha=-1$.  In fact, its value at $\alpha=-1$ is related to an important distance measure called quantum fidelity.

From the definition (\ref{Renyirel1}) it is clear that sandwich R{\'e}nyi divergences are larger than their cousins:\footnote{There is a potential for confusion here due to the notation we have chosen. It is sometimes said that Petz divergences are larger than sandwiched ones. That statement uses a different definition of Petz divergences which in our notation becomes the inequality (\ref{upperlower}) shown in appendix \ref{appA}. See section \ref{sec:finite}.}
\bea
S_\alpha(\Phi\|\Om)\geq D_\alpha(\Phi\|\Om)\ .
\eea

An analogy with finite dimensional quantum systems helps to unpack the definition in (\ref{Renyirel1}). 
This definition is based on a generalization of the notion of the $p$-norm of a matrix to unbounded operators \cite{araki1982positive}. Remember that for a matrix $X$ the $p$-norm is defined to be $\|X\|_p=tr(|X|^p)^{1/p}$. It is invariant under unitary rotations, $\|X\|_p=\|U^\dagger X U\|_p$, and it satisfies the Holder inequality 
\bea
\|XY\|_1\leq \|X\|_p\|Y\|_q\nn
\eea
for any $p,q>1$ with $p^{-1}+q^{-1}=1$. By the definition of the norm $| tr(X C)|\leq \|X C\|_1$ for any operator $C$.
As a result, 
\bea\label{normsmatrices}
\sup_{\|C\|_q=1}|tr(XC)|\leq \sup_{\|C\|_q=1}\|X C\|_1\leq \|X\|_p
\eea
where in the second step we have used Holder's inequality. If we pick $C$ proportional to $|X|^{p/q}U^\dagger$ where $X=U|X|$ is the polar decomposition of $X$ and the overall factor of $C$ is fixed such that $\|C\|_q=1$, we find that the inequality above is saturated. In other words, we have the norm duality relation
\bea\label{normdualmat}
\|X\|_p=\sup_{\|C\|_q=1}|tr(X C)|\ .
\eea
Our operators of interest are not trace class, however, choosing a reference state $|\Om\ra$ and an algebra of region $A$ following Araki and Masuda \cite{araki1982positive} one can still define the $p$-norm of a vector $|\Phi\ra$ using the definition
\bea\label{pnormgen}
&&p\in[2,\infty]\qquad \||\Phi\ra\|^A_{p,\Omega}=\sup_{|\Psi\ra}\|\Delta_{\Psi|\Om}^{1/2-1/p}|\Phi\ra\|\nn\\
&&p\in[1,2)\qquad \||\Phi\ra\|^A_{p,\Omega}=\inf_{|\Psi\ra}\|\Delta_{\Psi|\Om}^{1/2-1/p}|\Phi\ra\|\ .
\eea
As we discussed in section \ref{sec:modular}, we know that $(\Delta^A_{\Om|\Psi})^{-1}=\Delta^{A'}_{\Psi|\Om}$, therefore the sandwiched R{\'e}nyi divergence can be written as
\bea
S^A_\alpha(\Phi\|\Om)=\frac{2}{\alpha}\log\||\Phi\ra\|^{A'}_{\frac{2}{1-\alpha},\Om}\ .
\eea
The reason for switching from supremum to infimum as we go from positive to negative values of $\alpha$ is a generalization of (\ref{normdualmat}) to the norms defined in (\ref{pnormgen}) that relates the $p$ and $q$ norms when $\frac{1}{p}+\frac{1}{q}=1$. To see this, consider the Cauchy-Schwarz inequality 
\bea
\frac{|\la\Phi|\chi\ra|}{\la \Phi|\Delta_{\Om|\Psi}^{-\alpha}|\Phi\ra^{1/2}}\leq \la \chi|\Delta_{\Om|\Psi}^\alpha|\chi\ra^{1/2}\ .
\eea
Taking the infimum over $|\Psi\ra$ we have
\bea
\inf_{|\Psi\ra}\frac{|\la\Phi|\chi\ra|}{\la \Phi|\Delta_{\Om|\Psi}^{-\alpha}|\Phi\ra^{1/2}}\leq \inf_{|\Psi\ra}\la \chi|\Delta_{\Om|\Psi}^\alpha|\chi\ra^{1/2}\nn\ .
\eea
In fact, the supremum of the left-hand-side over all vectors $|\Phi\ra$ saturates the inequality \cite{araki1982positive}
\bea
\sup_{|\Phi\ra} \frac{|\la\Phi|\chi\ra|}{\sup_{|\Psi\ra}\la \Phi|\Delta_{\Om|\Psi}^{-\alpha}|\Phi\ra^{1/2}}= \inf_{|\Psi\ra}\la \chi|\Delta_{\Om|\Psi}^\alpha|\chi\ra^{1/2}\ .
\eea
As a result, we have the norm duality relation
\bea\label{normdual}
\sup_{\||\Phi\ra\|_{p,\Om}=1}|\la\Phi|\chi\ra|=\||\chi\ra\|_{q,\Om}\ .
\eea
This is a generalization of (\ref{normdualmat}) to von Neumann algebras.  
 
For special values of $\alpha$ the sandwiched R{\'e}nyi divergences in (\ref{Renyirel1}) is of particular interest in information theory:

\begin{itemize}

\item $\alpha=1$: 
Due to the monotonicity in $\alpha$, this is the maximum of the sandwiched R{\'e}nyi divergence in $\alpha$ and it is given by the norm of the operator that creates the state. If $|\Phi\ra=\Phi'|\Om\ra$ we have
\bea
S_1(\Phi\|\Om)&=&\sup_{|\Psi\ra}\log\la \Phi|\Delta_{\Om|\Psi}^{-1}|\Phi\ra=\sup_{|\Psi\ra}\log\la \Phi|S_{\Psi|\Om}S_{\Psi|\Om}^\dagger|\Phi\ra\nn\\
&=&\sup_{|\Psi\ra}\log\la \Psi|\|\Phi'\|^2|\Psi\ra=2\log\|\Phi'\|\nn
\eea
where we have used $(S^A_{\Psi|\Om})^\dagger=S^{A'}_{\Psi|\Om}$. An important implication of this result is that for states $\Phi'|\Om\ra$ with $\Phi'$ bounded all the sandwiched Renyi divergences are finite and free of ultraviolet divergences. This is a dense set of states.\footnote{Note that the R{\'e}nyi divergences including the relative entropy are not continuous. Therefore, in general, there exist states for which our measure is infinite. This is clear from the fact that the domain of $\Delta_{\Omega|\Psi}^{-1}$ is not the whole Hilbert space for all $|\Psi\ra$.}
Haag and Swieca have argued that the norm of the operator that creates the state should be interpreted as a measure of how localized the state is \cite{haag1965does}. Smaller $S_1(\Phi\|\Om)$ corresponds to a state more localized in $A'$. A unitary $U'$ creates an excitation completely localized in $A'$. 
 
\item $\alpha=\frac{1}{2}$: This value corresponds to a generalization of ``collision entropy" to von Neumann algebras: $S_{\frac{1}{2}}(\Phi\|\Om)=2\sup_\Psi\la\Phi|\Delta_{\Om|\Psi}^{-1/2}|\Phi\ra$. As we will see in the next section, this value of $\alpha$ can be related to Euclidean four point functions.

\item $\alpha=0$: Similar to the Petz divergence the sandwiched divergences is equal to the relative entropy at $\alpha=0$. To see this, we show the following inequality in appendix \ref{AppB}:
\bea\label{upperlower}
D_{\frac{\alpha}{1-\alpha}}(\Phi\|\Om)\geq S_\alpha(\Phi\|\Om)\geq D_\alpha(\Phi\|\Om)\ .
\eea
In the limit $\alpha\to 0$ the sandwiched R{\'e}nyi divergence is upper and lower bounded by relative entropy.

%
%
\item $\alpha=-1$: This value of $\alpha$ provides a generalization of ``quantum Fidelity" to von Neumann algebras:
\bea
&&S_{-1}(\Phi\|\Om)=-2\log F(\Phi,\Om)\nn\\
&&F(\Phi,\Om)\equiv \||\Phi\ra\|_{1,\Om}^{A'}
\eea
In the next section, we will show that for density matrices $F(\Phi,\Om)$ reduces to the standard expression for quantum Fidelity. From norm duality we know that Fidelity as defined above is the supremum over overlaps $\la \Psi|\Phi\ra$ for $|\Psi\ra=\Psi'|\Om\ra$ and $\|\Psi'\|=1$:
\bea
&&F(\Phi,\Om)=\sup_{\||\chi\ra\|_{\infty,\Om}^{A'}=1}|\la\chi|\Phi\ra|\ .
\eea
This is the analogue of Uhlemann's theorem in quantum field theory \cite{uhlmann1976transition}.
\end{itemize}

\section{Finite-dimensional Quantum Systems}\label{sec:finite}

Up to here, the discussion of the relative modular operator and our information-theoretic measures applied to any quantum system from qubits  to quantum field theory.
To come in contact with the information theory literature, in this section, we write our measures for finite quantum systems  in terms of density matrices, and 
provide a simple proof of the monotonicity of the Petz divergence and the sandwiched R{\'e}nyi divergence. We closely follow the arguments presented in \cite{nielsen2004simple} and \cite{Witten:2018zxz}; see also \cite{wilde2018optimized} for a similar proof. Consider a bipartite quantum system described by a density matrix $\sigma_{12}$ and its reduced density matrix on the first subsystem $\sigma_1$. It is convenient to purify $\sigma_{12}$ in a pure state of a four-partite system
\bea
|\Omega\ra=\sum_{i,j=1}^d c_{ij}|i j,i j\ra\in\mathcal{H}
\eea
where $|ij\ra$ are the eigenvectors of $\sigma_{ij}$ and we have assumed that the local Hilbert at each site is $d$-dimensional. In our notation, we separate the indices corresponding to the Hilbert space of the system from those of the purification by a comma. The coefficients $c_{ij}$ are the Schmidt coefficients of $|\Omega\ra$ and the square root of eigenvalues of $\sigma_{12}$. For simplicity, we assume that $\sigma_{12}$ is full-rank which implies that all $c_{ij}>0$.  Similarly, we purify $\sigma_1$ in a bi-partite pure state
\bea
&&|\omega\ra=\sum_i d_i |i,i\ra\in\mathcal{K}\nn\\
&&d_i^2=\sum_j c^2_{ij}.
\eea

The local algebra on each site is the algebra of $d\times d$ complex matrices with its natural notion of $\dagger$ and norm. We consider the algebras $\mathcal{A}=\mathcal{A}_1$ and $\mathcal{B}=\mathcal{A}_1\otimes \mathcal{A}_2$ that are generated by $a_{ik}=|i\ra\la k|$ and $b_{ijkl}=|ij\ra\la kl|$, respectively. Acting on the state $|\Omega\ra$ with $b\in\mathcal{B}$ we can obtain arbitrary states $\mathcal{H}$. Similarly, any state in $\mathcal{K}$ can be written as $a|\omega\ra$ with $a\in\mathcal{A}$.

The state $|\omega\ra$ can be though of as
\bea
&&|\omega\ra=\sigma_1^{1/2}|E_\omega\ra\nn\\
&&|E_\omega\ra=\sum_j |j,j\ra
\eea
where $|E_\omega\ra$ is an unnormalized maximally entangled state in the basis spanned by $|j\ra$, the eigenvectors of $\sigma_1$ \cite{Lashkari:2018oke}.
Note that for any full rank density matrix $\sigma$ we have
\bea\label{traceE}
\la \omega|X \otimes Y|\omega\ra=\la E_\omega|\sigma^{1/2}X \sigma^{1/2}\otimes Y|E_\omega\ra=tr( \sigma^{1/2}X \sigma^{1/2}Y)\ .
\eea
Consider the state $|\omega\ra$ on systems $13$ tensored with another copy of it on $24$: $|\omega\ra_{13}\otimes |\omega\ra_{24}$. If we act on it by the operator 
\bea
\tilde{W}_{34}=\sigma_{34}^{1/2} \lb \sigma_3^{-1/2}\otimes \sigma_4^{-1/2}\rb
\eea
we obtain the four-partite state $|\Omega\ra$:
\bea
|\Om\ra=\tilde{W}_{34} \lb |\omega\ra_{13}\otimes  |\omega\ra_{24}\rb 
\eea
where by $\sigma_{34}$ we mean $\sigma_{12}$ as a density matrix of system 34. 
Furthermore, we find that for $a\in\mathcal{A}_1$:
\bea
a|\Omega\ra=\tilde{W} \lb a|\omega\ra\otimes   |\omega\ra\rb 
\eea
Since $a|\omega\ra$ is dense in $\mathcal{K}$ we have constructed a linear linear embedding $W:\mathcal{K}\to \mathcal{H}$ that sends $a_{ik}|\omega\ra\to a_{ik}|\Omega\ra$. More explicitly, this map is
\bea
W=\sum_{ijk} \frac{c_{kj}}{d_k}|i j,k j\ra\la i,k|\ .
\eea

Consider the normalized pure states of subsystem $2$ and $4$: 
\bea
|\Psi_k\ra_{24}=\frac{1}{d_k}\sum_j c_{kj}|j,j\ra_{24}\ .
\eea
It is straightforward to check that
\bea\label{projector}
&&W^\dagger W=\mathbb{I}_\mathcal{K},\qquad W W^\dagger=\mathbb{I}_1\otimes \sum_k |k\ra_3\la k|_3\otimes |\Psi_k\ra_{24}\la \Psi_k|_{24}=P\nn
\eea
where $P$ is a projection in $\mathcal{H}$. Therefore, $W$ is an isometry. 

The anti-linear Tomita operator for algebras $\mathcal{A}$ and $\mathcal{B}$ satisfies
\bea
&&S_\omega^A a_{ik}|\omega\ra=a_{ik}^\dagger|\omega\ra\nn\\
&&S_\Omega^B b_{ijkl}|\Om\ra=b_{ijkl}^\dagger|\Om\ra\nn
\eea
which are solved by
\bea
&&S_\omega^A |k,i\ra=\frac{d_{k}}{d_{i}}|i,k\ra\nn\\
&&S_\Omega^B|kl,ij\ra=\frac{c_{kl}}{c_{ij}}|ij,kl\ra\ .
\eea
In this basis, the Tomita operator is the anti-linear operator that swaps the role of the system and its purifier. One can  compute the modular operators 
\bea\label{moddensity}
&&\Delta_\omega^A=\sum_{ik} \frac{d_k^2}{d_i^2}|k,i\ra\la k,i|=\sigma_1\otimes \sigma_3^{-1}\nn\\
&&\Delta_\Omega^B=\sum_{ijkl} \frac{c_{kl}^2}{c_{ij}^2}|k l,i j\ra\la kl ,ij|=\sigma_{12}\otimes \sigma_{34}^{-1}
\eea
and check directly that
\bea
W^\dagger |\alpha\beta,kj\ra\la \alpha\beta,kj|W= \frac{c_{kj}^2}{d_k^2}|\alpha,k\ra\la \alpha,k|
\eea
and therefore
\bea
&&\Delta_\omega^A=W^\dagger\Delta^B_\Omega W\ .
\eea
Since projectors are positive operators, for positive $\lam$ and $\beta$, we have
\bea
\lb \Delta^B_\Om+\beta\rb^{-1}&\geq& \lb \Delta^B_\Om+\beta+\lam(1-P)\rb^{-1}
\eea
At large $\lam$ the right-hand-side can be expanded as 
\bea
 \lb \Delta^B_\Om+\beta+\lam(1-P)\rb^{-1}&=& (P (\Delta^B_\Om+\beta) P)^{-1}+O(1/\lam)\nn\\
 &=&\lb  W (\Delta^A_\omega+\beta)W^\dagger \rb^{-1}+O(1/\lam)
\eea
where we have used $WW^\dagger=P$. Taking the $\lam\to\infty$ limit we find
\bea
W^\dagger(\beta+\Delta^B_\Omega)^{-1}W\geq (\beta+\Delta^A_\omega)^{-1}
\eea
as an operator inequality in $\mathcal{K}$. For a positive operator $\Delta$ and $0< \alpha <1$ we have the following spectral integrals 
\bea
&&\Delta^\alpha=\frac{\sin(\pi \alpha)}{\pi}\int_0^\infty d\beta\: \beta^\alpha\lb \frac{1}{\beta}-\frac{1}{\beta+\Delta}\rb\nn\\
&&\Delta^{-\alpha}=\frac{\sin(\pi \alpha)}{\pi}\int_0^\infty d\beta\:\beta^{-\alpha}\lb \frac{1}{\beta+\Delta}\rb\ .
\eea
Therefore, for $-1<\alpha<1$ we find
\bea
\sign(\alpha) \lb\Delta^A_\omega\rb^{-\alpha}\leq \sign(\alpha) \lb\Delta^B_\Omega\rb^{-\alpha}\ .
\eea

Now, consider a second density matrix $\phi_{12}$ and its purification
\bea
&&\phi_{12}=\sum_{\alpha\beta} f^2_{\alpha\beta}|\alpha\beta\ra\la\alpha\beta|\nn\\
&&|\Phi\ra=\sum_{\alpha\beta} f_{\alpha\beta} |\alpha\beta,\alpha\beta\ra\in\mathcal{H}\ .
\eea
Similarly, the density matrix $\phi_1$ can be purified with
\bea
&&|\varphi\ra=\sum_{\alpha} g_{\alpha} |\alpha,\alpha\ra\in\mathcal{K}\nn\\
&&g_\alpha^2=\sum_\beta f_{\alpha\beta}^2\ .
\eea
The relative Tomita operator now solves
\bea
&&S^A_{\varphi|\omega} a_{\alpha k}|\omega\ra= a^\dagger_{\alpha k}|\varphi\ra\nn\\
&&S^B_{\Phi|\Omega} b_{\alpha\beta kl}|\Omega\ra= b^\dagger_{\alpha\beta kl}|\Phi\ra
\eea
where $b_{\alpha\beta kl}=|\alpha\beta\ra\la kl|$ and $a_{\alpha k}=|\alpha\ra\la k|$. These equations imply
\bea
&&S^A_{\varphi|\omega}|\alpha,i\ra=\frac{g_\alpha}{d_i}|i,\alpha\ra\nn\\
&&S^B_{\Phi|\Omega} |\alpha\beta,ij\ra=\frac{f_{\alpha\beta}}{c_{ij}}|ij,\alpha\beta\ra\ .
\eea
The relative modular operators can be worked out explicitly:
\bea\label{relmodfinite}
&&\Delta_{\varphi|\omega}^A=\sum_{i\alpha} \frac{g_\alpha^2}{d_i^2}|\alpha,i\ra\la \alpha,i|=\phi_3\otimes \sigma_1^{-1}\nn\\
&&\Delta_{\Phi|\Omega}^B=\sum_{ij\alpha\beta} \frac{f_{\alpha\beta}^2}{c_{ij}^2}|\alpha\beta,ij\ra\la \alpha\beta,ij|=\phi_{12}\otimes \sigma_{34}^{-1}
\eea
From (\ref{traceE}) and (\ref{relmodfinite}) we find that the Petz divergence written in terms of density matrices is\footnote{Often the Petz divergence is defined in the range $\alpha\in (0,2)$ using $\bar{D}_n=\frac{1}{n-1}\log tr(\phi^n \omega^{1-n})$. In our notation, this is $D_{n-1}$.}
 \bea\label{Renyimatrices}
 &&D_\alpha(\Phi\|\Om)=\frac{1}{\alpha}\log tr\lb \phi^{1+\alpha}\sigma^{-\alpha}\rb\ .
 \eea
Once again, we can check explicitly that
\bea
W^\dagger\Delta^B_{\Phi|\Omega}W=\Delta_{\varphi|\omega}^A\ .
\eea
By the same argument we used for the modular operator, we find the monotonicity inequality
\bea\label{ineq}
 W^\dagger\lb\Delta_{\Phi|\Omega}^B+\beta\rb^{-1} W\geq \lb\Delta^A_{\varphi|\omega}+\beta\rb^{-1}\ .
\eea
Therefore, in the range $\alpha\in(-1,1)$ we have 
\bea\label{operatorinequal}
\sign(\alpha)(\Delta_{\varphi|\omega}^A)^{-\alpha}\leq \sign(\alpha)\: W^\dagger (\Delta_{\Phi|\Omega}^B)^{-\alpha} W
\eea
Evaluating the operator inequality above in state $|\omega\ra$, and using the fact that $W|\omega\ra=|\Omega\ra$ we get
\bea
\sign(\alpha)\la \omega| (\Delta_{\varphi|\omega}^A)^{-\alpha}|\omega\ra \leq \sign(\alpha) \la \Om| (\Delta_{\Phi|\Om}^B)^{-\alpha}|\Om\ra\ .
\eea
As a result, we find the monotonicity of the Petz divergence
\bea
D_\alpha(\omega\|\phi)\leq D_\alpha(\Om\|\Phi)\ .
\eea

Next, we would like to find the expression for the sandwiched R{\'e}nyi divergence in terms of density matrices and prove its monotonicity.
Consider a third vector $|\Psi\ra$ and its corresponding density matrix $\psi_{12}$, and $\psi_1$ with its purification $|\psi\ra$. 
Let us fix the region for the moment. It follows from the definition in (\ref{Renyirel1}) and the expressions in (\ref{relmodfinite}) that
 \bea
 &&S_{\alpha}(\Phi\|\Om)=\frac{1}{\alpha}\sup_{tr(\psi)=1}\log tr\lb \psi^{\alpha}\phi^{1/2}\sigma^{-\alpha}\phi^{1/2}\rb,
 \eea
 where we have used the cyclicity of the trace, and the supremum is over all density matrices $\psi$. 
 Defining the positive matrix $\eta=\psi^\alpha$ and $X=\sigma^{-\alpha/2}\phi^{1/2}$ we can write the sandwiched R{\'e}nyi divergence in (\ref{Renyimatrices}) in terms of $p$-norms as
\bea
&&S_{\alpha}(\Phi\|\Om)=\frac{1}{\alpha}\log\lb \sup_{\|\eta\|_{1/\alpha}=1}tr\lb \eta X^\dagger X\rb\rb
\eea
From (\ref{normdualmat}) the supremum can be evaluated explicitly to give
\bea\label{renyidisc}
S_{\alpha}(\Phi\|\Om)&=&\frac{1}{\alpha}\log\|X^\dagger X\|_{\frac{1}{1-\alpha}}=\frac{1}{\alpha}\log\|X X^\dagger\|_{\frac{1}{1-\alpha}}\nn\\
&=&\frac{1-\alpha}{\alpha}\log tr\left[\lb \sigma^{-\alpha/2}\phi\sigma^{-\alpha/2}\rb^{\frac{1}{1-\alpha}}\right] 
\eea
where we have used the polar decomposition of $X=U|X|$ and the fact that $p$-norm is unitary-invariant. In the literature, the sandwiched R{\'e}nyi divergence is often defined for $n\in(1/2,\infty)$ to be $\bar{S}_n=\frac{1}{n-1}\log tr\left[\lb \sigma^{\frac{1-n}{2n}}\rho\sigma^{\frac{1-n}{2n}}\rb^{n}\right]$. In our notation, this is $S_{ \frac{n-1}{n}}$. The statement that $\bar{D}_n\geq \bar{S}_n$ in our notation becomes $D_{n-1}\geq S_{\frac{n-1}{n}}$ which is the same as (\ref{upperlower}).
The expression in (\ref{renyidisc}) continues to hold when $\alpha<0$. 
This can be seen by computing the generalized $p$-norm in (\ref{pnormgen}) in terms of density matrices:
\bea
&&\||\Phi\ra\|^{A'}_{\frac{2}{1-\alpha},\Om}=\sup_{tr(\chi)=1}tr\lb \psi^{\alpha}\phi^{1/2}\sigma^{-\alpha}\phi^{1/2}\rb\nn\\
&&=tr_{\|\eta\|_{1/\alpha}=1}\lb \eta X^\dagger X\rb=\|X^\dagger X\|_{\frac{1}{1-\alpha}}=\|X\|_{\frac{2}{1-\alpha}}
\eea
where $X=\omega^{-\alpha/2}\phi^{1/2}$ and $\eta=\psi^{\alpha}$. Therefore,
\bea
S_{-\alpha}(\Phi\|\Om)&=&-\frac{1}{\alpha}\log \sup_{\|\sigma^{-\alpha/2}\psi^{1/2}\|_{\frac{2}{1-\alpha}}=1}|tr\lb \psi^{1/2}\phi^{1/2}\rb |\nn\\
&=&-\frac{1}{\alpha}\log\sup_{\|\nu\|_{\frac{2}{1-\alpha}}=1} |tr\lb \nu \phi^{1/2}\sigma^{\alpha/2}\rb |=-\frac{1}{\alpha}\log\|\phi^{1/2}\sigma^{\alpha/2}\|_{\frac{2}{1+\alpha}}\nn\\
&=&-\frac{1+\alpha}{\alpha}\log tr\left[\lb \sigma^{\alpha/2}\phi\sigma^{\alpha/2}\rb^{\frac{1}{1+\alpha}}\right] 
\eea
which is the same as (\ref{renyidisc}). 

Now, we are ready to prove the monotonicity of the sandwiched Renyi divergence. 
Assume $|\varphi\ra$ is cyclic and separating. Consider the subset of vectors in $\mathcal{H}$: $|\Psi_a\ra=a|\Phi\ra$ with $a$ an invertible element of $\mathcal{A}_1$, the algebra of the first subsystem. Their corresponding states on $\mathcal{K}$ are $|\psi_a\ra=a|\varphi\ra$ that form a dense set in $\mathcal{K}$ \cite{Lashkari:2018oke}.
The function $\la\varphi| \Delta_{\omega|\psi}^{-\alpha}|\varphi\ra$ is real-valued and continuous in finite-dimensional Hilbert spaces. Therefore, since $a|\psi\ra$ with invertible $a$ is dense in $\mathcal{K}$ we find
\bea
\sup_\psi \la\varphi|\lb\Delta^A_{\omega|\psi}\rb^{-\alpha}|\varphi\ra=\sup_{\psi_a}\la \varphi| \lb\Delta^A_{\omega|\psi_a}\rb^{-\alpha}|\varphi\ra\ .
\eea
From the operator inequality in (\ref{operatorinequal}) we have
\bea
\sign(\alpha)\sup_{\psi_a}\:\la \varphi| \lb\Delta^A_{\omega|\psi_a}\rb^{-\alpha}|\varphi\ra\leq \sign(\alpha)\sup_{\Psi_a}\:\la \varphi| W_{\Psi_a}^\dagger \lb\Delta^A_{\Omega|\Psi_a}\rb^{-\alpha}W_{\Psi_a}|\varphi\ra
\eea
We observe that 
\bea
W_{\Psi_a}|\varphi\ra= W_{\Psi_a}a^{-1}|\psi_a\ra=a^{-1}|\Psi_a\ra=|\Phi\ra\ .
\eea
Therefore,
\bea\label{monodensity}
&&\sign(\alpha)\sup_\psi \la\varphi|\lb\Delta^A_{\omega|\psi}\rb^{-\alpha}|\varphi\ra\leq \sign(\alpha)\sup_{\Psi_a}\la \Phi| \lb\Delta^A_{\Omega|\Psi_a}\rb^{-\alpha}|\Phi\ra\nn\\
&&\leq \sign(\alpha) \sup_{\Psi}\la \Phi| \lb\Delta^A_{\Omega|\Psi}\rb^{-\alpha}  |\Phi\ra
\eea
Hence, we have established a simple proof of the monotonicity of the sandwiched R{\'e}nyi divergence:
\bea
S_\alpha(\varphi\|\omega)\leq S_\alpha(\Phi\|\Omega)\ .
\eea

\section{Constraining Correlation Functions}\label{sec:constraint}

In this section, we start with the expression for the sandwiched R{\'e}nyi divergence in terms of density matrices in (\ref{renyidisc}) and use the Euclidean path-integral formalism to write them in terms of correlation functions. At the first look, this appears to be against our philosophy to avoid the use of density matrices in quantum field theory. However, since we defined our R{\'e}nyi families directly within the modular theory we are guaranteed that they are ultraviolet finite for states $\Phi|\Om\ra$ with $\Phi$ an operator in the algebra or the commutant. This is manifestly the case in \cite{lashkari2014relative} that writes the sandwiched R{\'e}nyi divergence for values of $\alpha=1-\frac{1}{n}$ and positive integer $n>1$ in terms one-sheeted Euclidean $2n$-point correlation functions. We review this construction below. It is to be contrasted with the standard expression for R{\'e}nyi entropies that is an $n$-sheeted partition function and ultraviolet divergent. Unfortunately, the Petz divergence involves fractional powers of both density matrices for all values of $\alpha$ and we cannot have a simple Euclidean path-integral representation for it, except for the case that the excited state is the vacuum of the theory deformed with a relevant operator, e.g. see \cite{bernamonti2018holographic}. For the remainder of this section, we focus on the sandwiched R{\'e}nyi divergence.

\subsection*{Path-integral approach}
\begin{figure}
\centering
\includegraphics[width=\textwidth]{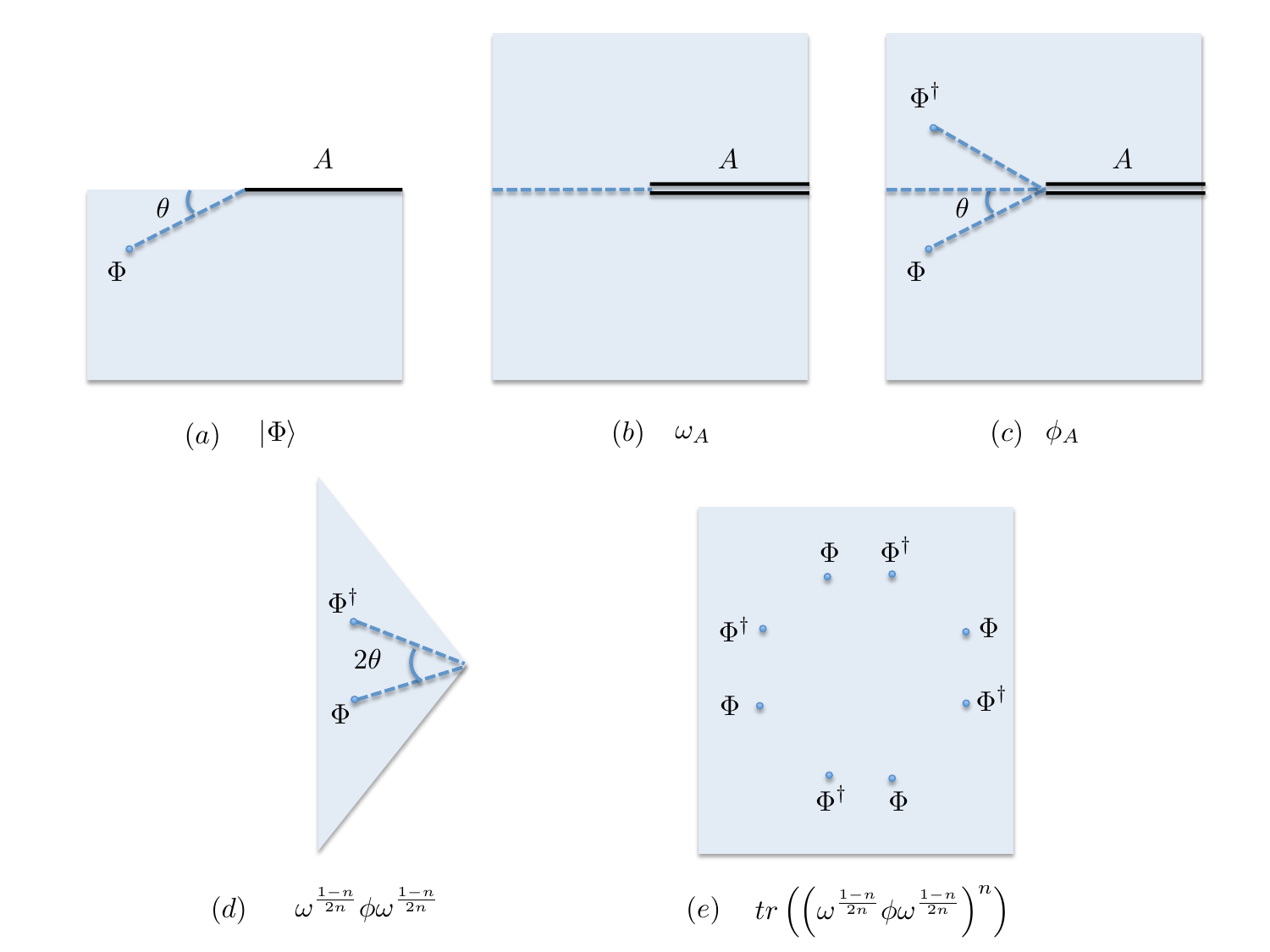}\\
\caption{\small{The Euclidean path integrals that prepare (a) excited state $|\Phi\ra$, (b) vacuum density matrix on region $A$, (c) the density matrix of $A$ in the excited state. (d) The operator $\omega^{\frac{1-n}{2n}}\phi\omega^{\frac{1-n}{2n}}$ corresponds to a path-integral on a wedge of angular size $2\pi/n$. (e) The correlator that appears in the definition of the sandwiched R{\'e}nyi divergence.}}
\label{fig1}
\end{figure}

Consider the vacuum of a quantum field theory $|\Om\ra$, an operator $\Phi$ supported in $A'$ and excited states $|\Phi\ra=\Delta^{-\theta/\pi} \Phi|\Om\ra$ with  $0\leq \theta\leq 1/2$ \cite{Witten:2018zxz}. These states can be represented by a Euclidean path-integral in the lower half-plane with an operator $\Phi$ inserted at angle $-(\pi-\theta)$; see figure \ref{fig1}. 
The vacuum density matrix $\omega_A$ for half-space $x^1>0$ is the path-integral on the whole plane with a cut above and below the positive real line, and similarly for $\phi_A$ with two operator insertions $\Phi$ and $\Phi^\dagger$ at angle $-(\pi-\theta)$ and $(\pi-\theta)$, respectively.
If $\theta+\pi\alpha\leq \pi$ there is a path-integral representation for the operator $\omega^{-\alpha/2}\phi\omega^{-\alpha/2}$ as a wedge of angle $2\pi(1-\alpha)$ with two operator insertions; figure \ref{fig1}. If we choose $\alpha$ such that $1-\alpha=\frac{1}{n}$ with $n>1$ a positive integer, $n$ such wedges can be sewn together to obtain $tr\lb (\omega^{\frac{1-n}{2n}}\phi\omega^{\frac{1-n}{2n}})^{n}\rb$  as a $2n$-point function of $\Phi$ and $\Phi^\dagger$ with operators inserted at $z^{(\pm)}_k=r e^{i(\frac{2k\pi}{n}\pm\theta)}$ for $k=0,\cdots, n-1$:
\bea
tr\lb (\omega^{\frac{1-n}{2n}}\phi\omega^{\frac{1-n}{2n}})^{n}\rb=\left\la \prod_{k=0}^{n-1} \Phi^\dagger(z^+_k)\Phi(z^-_k)\right\ra
\eea
where we have suppressed the $D-2$ perpendicular directions $x_\perp$.
 For integer $n>1$ the sandwiched R{\'e}nyi divergence in (\ref{renyidisc}) is given by the following Euclidean correlation function
\bea\label{renyi2n}
S^A_{1-\frac{1}{n}}(\Phi\|\Om)=\frac{1}{n-1}\log\lb\frac{\left\la\prod_{k=0}^{n-1}\Phi^\dagger(z^+_k)\Phi(z^-_k)\right\ra}{\la \Phi^\dagger(z^+_0)\Phi(z^-_0)\ra^n} \rb\ .
\eea
The allows us to relate the monotonicity of sandwiched R{\'e}nyi divergences to constraints on correlation functions. It is worth noting that the non-negativity of sandwiched R{\'e}nyi divergence is already a non-trivial constraint implying that for all positive integer $n>1$ the numerator in (\ref{renyi2n}) is larger than the denominator.

We would like to understand how the correlator in (\ref{renyi2n})  changes as we change the region $A$. As we discussed in section \ref{sec:modular}, for a one-parameter family of smooth deformations given by $U_\lambda$ with $U_0=1$ that are symmetries of the vacuum state, the relative modular operator transforms according to 
\bea
\Delta_{\Psi|\Om}^{A(\lam)}=U_\lam^\dagger \Delta^A_{U_\lam\Psi|\Om} U_\lam\ .
\eea
Plugging this into the definition of the sandwiched R{\'e}nyi divergence we find
\bea\label{transdiff}
S^{A(\lam)}_\alpha(\Phi\|\Om)=S_\alpha^A(U_\lam \Phi\|\Om)\ .
\eea

The unitary $U_\lambda$ acts geometrically on local operators: $\Phi(z)\to \Phi^{(\lambda)}(z_\lambda)$. The sandwiched R{\'e}nyi divergence of the deformed region $A(\lambda)$ is given by the expression in (\ref{renyi2n}) with $\Phi(z)$ replaced with $\Phi^{(\lambda)}(z_\lambda)$.

To be specific, we consider the Rindler space parameterized by $(z,x_\perp)$ where $z=x+i \tau$ and $x_\perp$ are the perpendicular directions. We focus on two classes of shape deformations and assume that operator $\Phi$ is a scalar:
\begin{enumerate}
\item {\bf Translations:} The unitary $U_\lam(\zeta)=e^{i\lam P_\zeta}$ with $P_\zeta=\zeta_0 P^0+\zeta_1 P^1$ corresponds to a translation in $\zeta$ direction.

\item {\bf Null deformations:} The unitary $U_\lam(f,x_\perp)=e^{i \lam Q_f}$ with $Q_f=\int du f(x_\perp) T_{uu}(u,x_\perp)$ a conserved charge associated with deformations on a null hypersurface parameterized by $u=t-r$ and $x_\perp$. 
\end{enumerate}
The second class includes the first one as a special case.\footnote{The statement that the second class of transformation are symmetries of the vacuum is the so-called Markov property of the vacuum in quantum field theory \cite{casini2017modular,lashkari2017entanglement}.} However, it is harder to work with $Q_f$ because, as opposed to $P_\zeta$, it is not a topological charge. If the states we consider are created by the action of only one operator acting on the vacuum, then $U_\lam(f,x_\perp)$ acts as:
\bea
U_\lam(f,x_\perp)\Phi(z,x_\perp)|\Om\ra=U_\lam(\zeta)\Phi(z,x_\perp)|\Om\ra,
\eea
where $\zeta=f(x_\perp)\p_u $ is a null translation.
 We will use this simplifying assumption and postpone the study of the consequences of monotonicity under null deformations for general states to future work.

Consider the Euclidean translation $(\tau,x_1)\to (\tau+\lambda \zeta_0,x_1+\lambda)$ then 
\bea\label{movingop}
&&\Phi(-\tau_0,x_1)\to e^{-\lambda (\zeta_0P^0+P^1)} \Phi(-\tau_0,x_1)e^{\lambda(\zeta_0P^0+P^1)}=\Phi(-\tau_0-\lambda \zeta_0,x_1-\lambda ),\nn\\
&&\Phi^\dagger(\tau_0,x_1)\to e^{-\lambda(-\bar{\zeta}_0P^0+P^1)} \Phi^\dagger(\tau_0,x_1)e^{\lambda(-\bar{\zeta}_0P^0+P^1)}=\Phi^\dagger(\tau_0+\lambda \bar{\zeta}_0,x_1-\lambda ),
\eea
where we have taken $a_0$ to be complex so that we can analytically continue it to real time. If we take $a_0$ to be pure imaginary the pair $\Phi^\dagger\Phi$ move together and the denominator in (\ref{renyi2n}) stay the same. 
The other pairs of operators $\Phi^\dagger\Phi$ in the numerator of (\ref{renyi2n}) are the same as $\Phi^\dagger\Phi$ above but rotated by $2\pi k/n$. To simplify our notation, we refer to $\Phi(z_k)$ as $\Phi_k$.
 
\begin{figure}
\centering
\includegraphics[width=\textwidth]{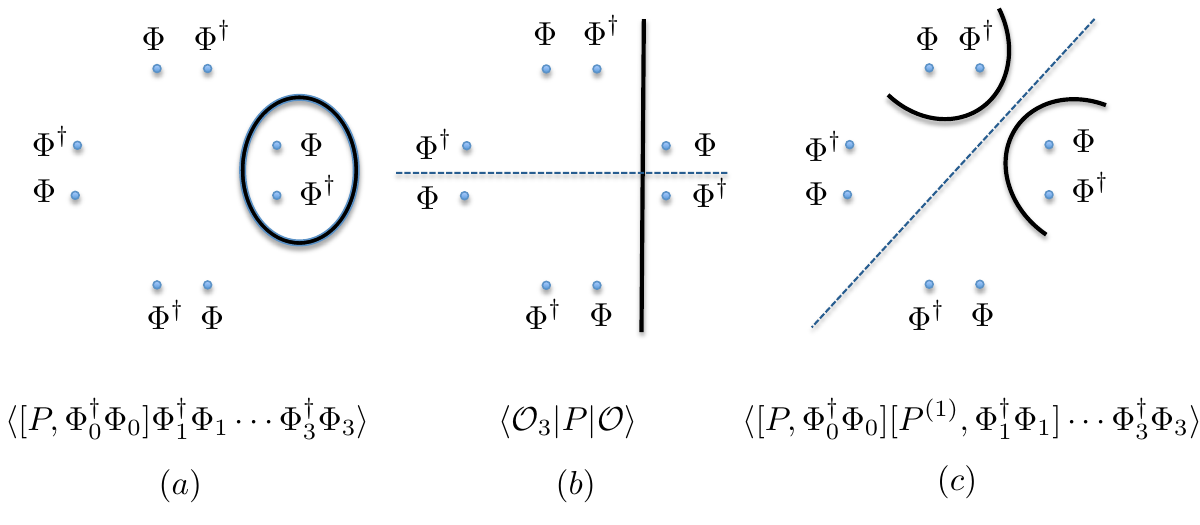}\\
\caption{\small{(a) The first derivative in shape deformation of the sandwiched R{\'e}nyi divergence for $n=4$. The commutator is represented with the charge written as an integral on the codimension one surface encircling the pair $\Phi_0^\dagger\Phi_0$. (b) Same commutator with the charges written on a different surface, (c) A term that appears at the second order in deformation which is reflection symmetric around $\theta=\pi/2$. Note that $P^{(1)}$ is the same as $P$ but rotated by $\pi/2$.}}
\label{fig2}
\end{figure}

  Changing the region corresponds to moving $\Phi$ in the path-integral. The first derivatives in shape deformation corresponds to the first $\lam$ derivatives in (\ref{movingop}). The $m^{th}$ $\lam$ derivative is a nested commutator of the pair $\Phi^\dagger\Phi$ with the momentum that generates the translation:
 \bea\label{nested}
\frac{d^m}{d\lambda^m}(\Phi^\dagger\Phi)=\frac{1}{m!}[P_\zeta,[P_\zeta,\cdots, [P_\zeta,\Phi^\dagger\Phi]]]\ .
\eea
The momentum $P_\zeta$ is a topological charge, so it can be written as 
\bea
P_\zeta=\int_C d\Sigma^i \zeta^j T_{ij},
\eea
where $C$ is any codimension one surface that encircles the pair $\Phi^\dagger\Phi$ and no other operators.
We choose this surface to be $x_1=c$ with $\tau_0\sin(\pi/n)<c<x_1$; see figure \ref{fig2}.  The momentum $P_\zeta$ written on this surface is
\bea
P_\zeta=\int d\tau \lb  T_{x_1x_1}+\zeta_0 T_{\tau x_1} \rb=P^1+\zeta_0 P^0,
\eea
where $P^0$ and $P^1$ are momenta in $\tau$ and $x_1$ directions, respectively. 
We are interested in deformations of region $A$ that send $(x_0, x_1)\to (x_0+i t, x_1+\lam)$ with $t<1$, so we set $\zeta_0=i t$.  This makes sure $A(\lam)\subset A(\lam')$ for any $\lam< \lam'$. 

To the first order in $\lambda$ by rotation symmetry we have 
\bea\label{highernfirst}
\p_\lambda S^{A(\lambda)}_{1-\frac{1}{n}}(\Phi\|\Om)_{\lambda=0}=\frac{n\la [P_\zeta,\Phi^\dagger_0\Phi_0 ]\Phi^\dagger_1\Phi_1\cdots \Phi^\dagger_{n-1}\Phi_{n-1}\ra}{\la\Phi^\dagger_0\Phi_0\cdots \Phi^\dagger_{n-1}\Phi_{n-1}\ra}
\eea
where the factor of $n$ comes the rotation invariance of correlators. It is convenient to interpret the correlator above by choosing $\p_{x_1}$ as the generator of time-translations. In this quantization frame, $P^1$ becomes the Hamiltonian and $P_L=iP^0$ becomes the Lorentzian momentum along a spacelike direction; see figure \ref{fig2}. 
Therefore, the monotonicity of R{\'e}nyi divergences can be understood as the non-negativity of the following off-diagonal matrix elements of $H-t P_L$ with $0<t<1$:
\bea\label{constraint}
\forall n\geq 1:\qquad \la \Phi_0^\dagger\Phi_0|H-t P_L|\Phi_1^\dagger\Phi_1\cdots \Phi_{n-1}^\dagger\Phi_{n-1}\ra\geq 0\ .
\eea
Note that the denominator in (\ref{highernfirst}) is manifestly positive from reflection positivity around $\tau=0$.
Since $H-t P_L=t P_u+(1-t)H$ with $P_u$ the null momentum in $u$ direction, we only need to consider the constraint at $t=0$ and $t=1$.
When $n=2$ the inequality above follows from the positivity of $P_u$. To our knowledge, for $n>2$ this is a new constraint that does not follow from any of known inequalities in a trivial way.\footnote{According to the reconstruction theorems the reflection positivity and the analyticity of Euclidean correlators are sufficient to insure unitarity and causality in Lorentzian signature \cite{osterwalder1975axioms}. This suggests that there has to be a way to derive this constraint from reflection positivity. However, we could not find such an argument for $n>2$.} The physical interpretation of this result is that not only $H$ and $P_u$ are positive operators but also they have the following non-zero off-diagonal matrix elements:
\bea
\la \mO| P_u|\mO_{n-1}\ra\geq 0,\qquad \la \mO| H|\mO_{n-1}\ra\geq 0
\eea
where $|\mO\ra= |\Phi_0^\dagger\Phi_0\ra$ and $|\mO_{n-1}\ra=|\Phi^\dagger_1\Phi_1\cdots \Phi^\dagger_{n-1}\Phi_{n-1}\ra$. 

Consider the quantization frame that chooses $\tau$ as the Euclidean time. Now the $2n$-point function without momentum commutators can be thought of as the the norm of a state $|\chi\ra$ where $|\chi\ra=|\Phi_0 \Phi_1^\dagger\Phi_1\cdots \Phi^\dagger_{n/2}\ra$ if $n$ is even and $|\chi\ra=|\Phi_0 \cdots \Phi_{(n-1)/2}^\dagger\Phi_{(n-1)/2}\ra$ if $n$ is odd.

In the special case $\Phi=\Phi^\dagger$ the correlator is symmetric under the reflection $\tau\to -\tau$. Since this reflection sends $P^0$ to $-P^0$ but preserves $P^1$ we find that
\bea
&&\p_tS_{1-\frac{1}{n}}(\Phi\|\Om)|_{t=0}=0\nn\\
&&\p_x S_{1-\frac{1}{n}}(\Phi\|\Om)|_{t=0}\geq 0\ .
\eea 
Clearly, this holds only for the first derivative and just in the special case of a real field $\Phi=\Phi^\dagger$.\footnote{It we tune initial state to have be a real field, as the operator evolves in time it becomes a complex field.} 

\subsection*{A conjecture}
The sandwiched R{\'e}nyi divergence is smooth at $\alpha=0$ and equal to the relative entropy. It was conjectured and recently argued that, in addition to the first derivative of relative entropy in deformation parameter that is positive due to monotonicity, the second null derivative of relative entropy is also non-negative in field theory \cite{bousso2016quantum,balakrishnan2017general}. This is called the quantum null energy condition. In this subsection, we investigate the higher derivatives of the sandwiched R{\'e}nyi divergence.


Let us start with $n=2$. From (\ref{nested}) we know that the $m^{th}$ derivatives in null deformations is the expectation value of the positive operator $P_u^m$ in the state $|\mO\ra$:
\bea
\p_\lam^mS_{1/2}(\Phi\|\Om)=\frac{1}{m!}\la\mO| P_u^m|\mO\ra\geq 0\ .
\eea
Next, consider the second space-like derivative $(t=0)$ of R{\'e}nyi divergences with $n>2$ for states that satisfy $\Phi=\Phi^\dagger$. Then,
\bea
\p^2_x S^{A(\lambda)}_{1-\frac{1}{n}}(\Phi\|\Om)_{\lambda=0}&=&\frac{n}{2}  \frac{\la [P^{1},[P^1,\Phi_0\Phi_0 ] ]\cdots \Phi_{n-1}\Phi_{n-1}\ra}{\la\Phi_0\Phi_0 \cdots \Phi_{n-1}\Phi_{n-1}\ra}\nn\\
&&+\frac{n}{2} \sum_{k=2}^{n-1}\frac{\la [P^{1},\Phi_0\Phi_0 ] \cdots [P^{(1,k)},\Phi_{k-1}\Phi_{k-1}]\cdots \Phi_{n-1}\Phi_{n-1}\ra}{\la\Phi_0\Phi_0 \cdots \Phi_{n-1}\Phi_{n-1}\ra}\nn
\eea
where $P^{(1,k)}$ is the same operator as $P^{(1)}$ rotated by $2\pi k/n$; see figure \ref{fig2}.
The first term above is symmetric under reflection around $\theta=0$, and the $k^{th}$ term in the sum is symmetric under the reflection around $\theta=\pi k/n$; see figure \ref{fig2}. Therefore, for the deformation generated by $\p_{x_1}$ the second derivative of R{\'e}nyi relative entropy is non-zero for all $n$ if $\Phi=\Phi^\dagger$:
\bea
\p_x^2S(\Phi\|\Om)\geq 0\ .
\eea
Repeating the argument above for a null deformation and the special state $\Phi=\Phi^\dagger$ we find that the second null derivative is also positive. Therefore, for this special class of states, we have proved our conjecture 
\bea
\p_u^2S(\Phi\|\Om)\geq 0\ .
\eea

\subsection*{Examples}
It is instructive to explicitly compute the sandwiched R{\'e}nyi divergence and its derivatives in some simple theories.\footnote{In a recent work the R{\'e}nyi divergences were computed in free field theory using real-time techniques \cite{casini2018renyi}. Our Euclidean approach has the advantage that it makes the monotonicity constraints manifest.}  The theory has to be simple enough that we can compute the $2n$-point correlation functions. Two instances when we can access these correlators are free fields and small $\theta$ limit in conformal field theory.\footnote{It would be interesting to study these constraints in large $N$ theories.} 

Consider a two dimensional massless boson and a coherent operator $e^{i\beta\phi}$ with real $\beta$. Since this operator is a conformal primary, and there are the same number of operators in the numerator and the denominator of (\ref{renyi2n}), the sandwiched R{\'e}nyi divergence is independent of $r$ in $z_k^{(\pm)}=r e^{i(\frac{2\pi k}{n}\pm \theta)}$.   
The $n$-point function of the coherent operator is given by \cite{francesco2012conformal}
\bea
\left\la\prod_{k=0}^{n-1}e^{-i\beta_k \phi}(z_k)\right\ra=\prod_{k>i}(z_k-z_i)^{\beta_i \beta_k}\ .
\eea
For the initial configuration at $\lam=0$ we have
\bea
&&\frac{\left\la\prod_{k=0}^{n-1}e^{-i\beta \phi}(z^+_k)e^{i\beta \phi}(z^-_k)\right\ra}{\left\la e^{-i\beta \phi}(z^+_0)e^{i\beta \phi}(z_0^-)\right\ra^n}=\prod_{m=1}^{n-m}\lb\frac{\sin^2\lb \frac{\pi m}{n}\rb}{\sin\lb \frac{\pi m}{n}+\theta\rb\sin\lb \frac{\pi m}{n}-\theta\rb} \rb^{(n-m)\beta^2}=\lb \frac{n\sin\theta}{\sin(n\theta)}\rb^{n \beta^2}\nn
\eea
The sandwiched R{\'e}nyi divergence is \cite{lashkari2014relative} 
\bea
S_n^{A(0)}(e^{i\beta \phi}\|\Om)=\frac{n \beta^2}{n-1}\log\lb \frac{n\sin\theta}{\sin(n \theta)}\rb\ .
\eea
Initially starting at $(\tau=r\sin\theta,x_1=r\sin\theta)$ and deforming the region, the operator $\Phi_0$ moves to $(-\tau_0(1+\lam \zeta_0), x_1(1-\lam))$ and $\Phi^\dagger$ goes to $(\tau_0(1+\lam \bar{\zeta}_0), x_1(1-\lam))$, where $\bar{\zeta}_0$ is the complex conjugate of $\zeta_0$. Choosing $\zeta_0=i t$ we find that $\la\Phi^\dagger_k\Phi_k\ra$ remains unchanged.
For displaced operators from the $n$-point function formula we have
\bea
&&\frac{\left\la\prod_{k=0}^{n-1}e^{-i\beta \phi}(z^+_k(\lam))e^{i\beta \phi}(z^-_k(\lam))\right\ra}{\left\la e^{-i\beta \phi}(z^+_0)e^{i\beta \phi}(z_0^-)\right\ra^n}=\prod_{m=1}^{n-1}\lb \frac{4 c(\theta)c(-\theta)\sin^4\lb\frac{m\pi}{n} \rb}{X(\theta)X(-\theta)}\rb^{(n-m)\frac{\beta^2}{2}}\nn\\
&&X(\theta)=1-\cos \left(\frac{2\pi  m}{n}+2\theta \right)+2\lambda  \left(\cos \left(\frac{2 \pi  m}{n}+\theta\right)-\cos (\theta )\right)\nn\\
&&+\lambda ^2(1-t^2)\lb 1-\cos \left(\frac{2 \pi  m}{n}\right)\rb\nn\\
&&c(\theta)=1-2\lam(\cos\theta+i t \sin\theta)+\lam^2(1-t^2)\ .
\eea

First consider the case $n=2$. Then,
\bea
&&X(\theta)=1+\cos(2\theta)-4\lam\cos\theta+2\lam^2(1-t^2)
\eea
and
\bea
S_{1/2}^{A(\lam)}(e^{i \beta\phi}\|\Om)&=&\frac{\beta^2}{2}\log \left(\frac{4 \left(4 \lambda ^2 t^2 \sin ^2(\theta )+\left(1-2 \lambda  \cos (\theta )+\lambda ^2 \left(1-t^2\right)\right)^2\right)}{\left(1+\cos (2 \theta )-4 \lambda  \cos (\theta )-2 \lambda ^2 \left(t^2-1\right)\right)^2}\right)\ .
\eea
At $t=0$ we have
\bea\label{secondrenyispace}
S_{1/2}^{A(\lam)}(e^{i \beta\phi}\|\Om)=\beta^2\log\lb \frac{(1+\lam^2-2\lam\cos\theta)}{(\lam-\cos\theta)^2}\rb
\eea
According to Bernstein's theorem a function $f(\lam)$ has $\p_\lam^k f(\lam)\geq 0$ for all $k$ if and only if $f(-\lam)$ is the Laplace transform of a probability distribution $\mu(s)$ \cite{bernstein1929fonctions}:
\bea
f(-\lam)=\int_0^\infty ds\: e^{-s\lam} \mu(s)\ .
\eea
In many physics applications, the probability distribution $\mu(s)$ corresponds to the density of states that is a probability measure, and $f(-\lam)$ is the partition function whose odd (even) derivatives in inverse temperature are negative (positive). The second R{\'e}nyi divergence has an inverse Laplace transform
\bea
&&S_{1/2}^{A(-\lam)}(e^{i \beta\phi}\|\Om)=\int_0^\infty ds\: e^{-\lam s}\mu(s)\nn\\
&&\mu(s)=2\beta^2 \frac{e^{-s\cos\theta}(1-\cos(s\sin\theta))}{s}\geq 0\nn\\
&&\int_0^\infty ds\: \mu(s)=2\beta^2\log\sec\theta<\infty
\eea
which is indeed a probability measure. Therefore, we learn that all spatial derivatives of the second R{\'e}nyi divergence are non-negative.
Next, we consider $t=1$ that corresponds to a null deformation:
\bea\label{nullsecondrenyi}
S_{1/2}(e^{i \beta\phi}\|\Om)=\frac{\beta^2}{2}\log\lb \frac{1+4\lam^2-4\lam\cos\theta}{(2\lam-\cos\theta)^2}\rb
+\beta^2\log\sec\theta
\eea
The $\lam$ dependent piece in (\ref{nullsecondrenyi}) is the same as (\ref{secondrenyispace}) with $\lam\to 2\lam$, hence all of its $\lam$ derivatives as was proved before.

Same logic generalizes to $n>2$. 
Writing sandwiched R{\'e}nyi for $n>2$ and $t=0$ as a Laplace transform we find
\bea
&&S_{1-\frac{1}{n}}^{A(-\lam)}(e^{i\beta \phi}\|\Om)=\frac{2\beta^2}{n-1}\sum_{m=1}^{n-1}(n-m)\int_0^\infty ds e^{-\lam s} \mu(s)\nn\\
&& \mu(s)=\frac{e^{-s\cos\theta}}{s}\:\lb \cosh\lb s\sin\theta\cot\lb \frac{m \pi}{n}\rb\rb-\cos(s\sin\theta)\rb\geq 0
\eea
where both the numerator and the denominator are manifestly positive for $0\leq \theta\leq \pi/n$ and integer $n$. In the null case, it is easier to consider 
\bea
&&S^{A(\lam)}_{1-\frac{1}{n}}(e^{i \beta\phi}\|\Om)-\frac{1}{2}S^{A(0)}_{1-\frac{1}{n}}(e^{i \beta\phi}\|\Om)=\frac{\beta^2}{(n-1)}\sum_{m=1}^{n-1}(n-m)\int_0^\infty ds\: e^{-s\lam} \mu(s)\nn\\
&&\mu(s)= \frac{e^{-\frac{s}{2}\cos\theta}}{s}\:\lb \cosh\lb \frac{s}{2}\sin\theta\cot\lb \frac{m \pi}{n}\rb\rb-\cos\lb\frac{s}{2}\sin\theta\rb\rb\geq 0\ .
\eea
To summarize, we established that all spatial and null derivatives of R{\'e}nyi divergences of coherent states with respect to vacuum are positive. 
The fact this holds for higher than two derivatives in deformations is special to coherent states. To show this, we consider the chiral primary operator $\p\phi$ of the same theory. 
\begin{figure}
\centerline{%
\includegraphics[width=0.5\textwidth]{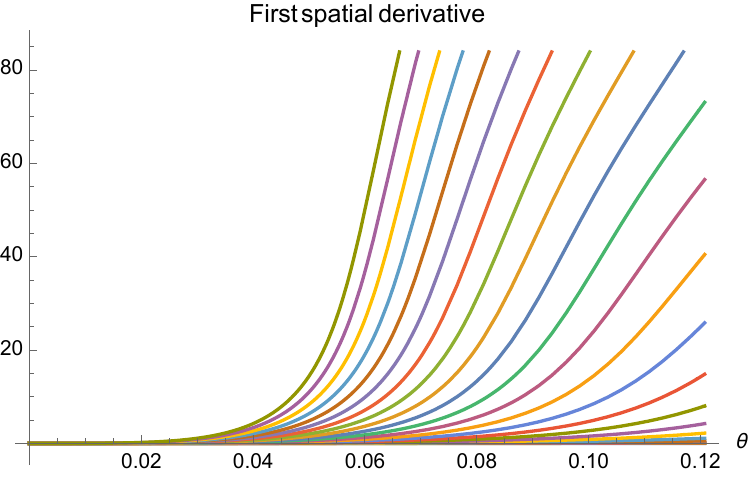}%
\includegraphics[width=0.5\textwidth]{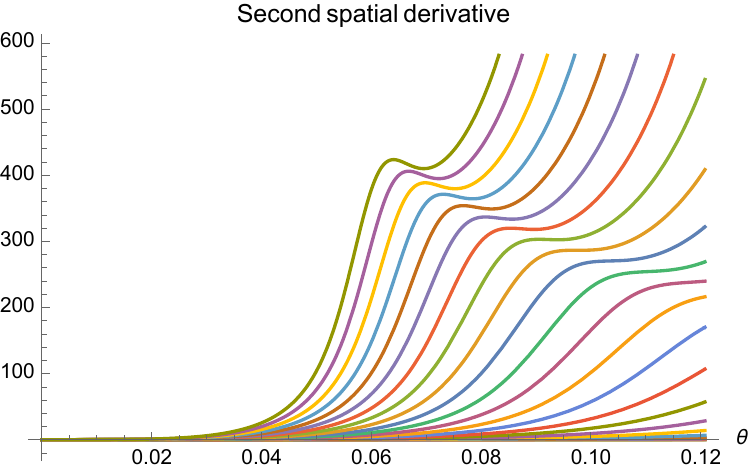}%
}%
\centerline{%
\includegraphics[width=0.5\textwidth] {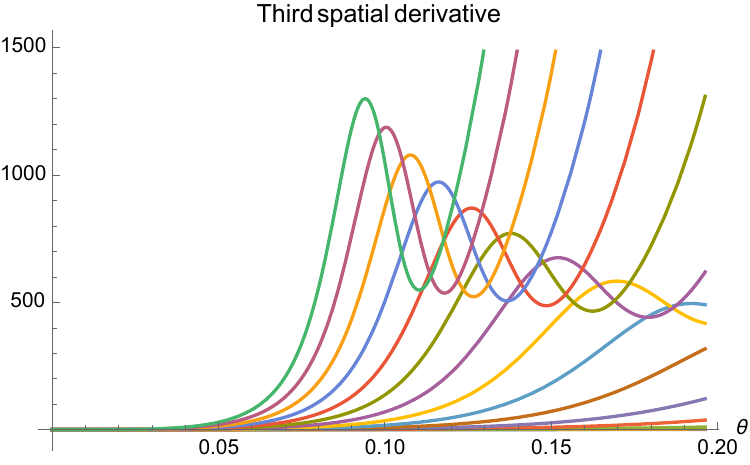}%
\includegraphics[width=0.5\textwidth] {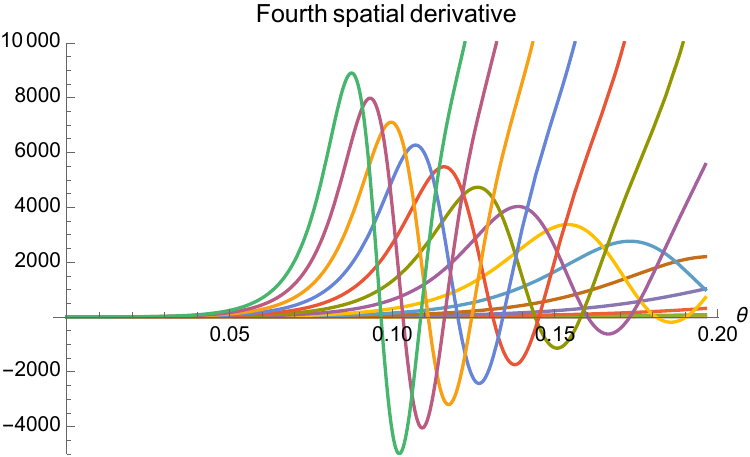}%
}%
\caption{\small{(top) The first and the second spatial derivatives of the sandwiched R{\'e}nyi divergence for $n=2$ to $n=26$ are non-negative and consistent with monotonicity and our proof. (bottom) The third and the fourth spatial derivatives are plotted for $n=2$ to $n=16$. For large enough $n$, the fourth derivative becomes negative.}}
\label{fig3}
\end{figure}
This operator has dimension $(1,0)$ and its $2n$-point functions were computed in \cite{calabrese2014entanglement}:
\bea
G_n=\left\la \prod_{i=0}^{2n}(\p\phi)^\dagger(\p\phi)\right\ra=\frac{1}{4^n}\det\left[ \frac{1}{\sin\lb\frac{z_i-z_j}{2}\rb}\right]_{i,j\in[1,2n]}=\frac{\Gamma^2\lb \frac{1+n+n\csc(n\theta)}{2}\rb}{\Gamma^2\lb \frac{1-n+n\csc(n\theta)}{2}\rb}\ .
\eea
Therefore,
\bea
S_{1-\frac{1}{n}}(\p\phi\|\Om)=\frac{1}{n-1}\log\lb (2\sin\theta)^{2n} G_n\rb
\eea
A spatial deformation correponds to $\cos\theta\to \cos\theta-\lam$ in the formula above. Then, the first spatial derivative is given by
\bea
&&\p_\lam S^{A(\lam)}_{1-\frac{1}{n}}(i\p\phi\|\Om)_{\lam=0}\nn\\
&&=\frac{2n\cos\theta}{(n-1)}\lb 1-\frac{n\tan\theta}{\tan(n\theta)\sin(n\theta)}\lb \psi\lb  \frac{1+n+n\csc(n\theta)}{2}\rb-\psi\lb \frac{1-n+n\csc(n\theta)}{2}\rb\rb\rb\nn\ .
\eea
Note that this operator is real, so our proof of the non-negativity of the second derivative in spatial deformations applies to it. Figure \ref{fig3} is a plot of the first four derivatives in spatial deformations. The first and the second derivatives of sandwiched R{\'e}nyi divergence are non-negative as expected, however the fourth derivative becomes negative for large enough $n$.


As the second example, we consider the states of an interacting conformal field theory with $\theta\ll 1$.
 We expand each pair of operators in small $\theta$ using the operator product expansion:
\bea
&&\frac{\Phi_k^\dagger\Phi_k}{\la\Phi_k^\dagger \Phi_k\ra}=\sum_p (2\sin\theta)^{h_p} C^p_{\phi\phi}\mO_p(z_k^0,z_k^1)\nn\\
&&z_k^0=(\cos\theta-\lam)\sin\lb \frac{2\pi m}{n}\rb-(\sin\theta+i t \lam)\cos\lb \frac{2\pi m}{n}\rb\nn\\
&&z_k^1=(\cos\theta-\lam)\cos\lb \frac{2\pi m}{n}\rb-(\sin\theta+i t \lam)\sin\lb \frac{2\pi m}{n}\rb,
\eea
where $\mO_p$ represent primaries and their descendants. The $2n$-point function in this limit is
\bea
&&\log\lb\frac{\left\la\prod_{k=0}^{n-1}\Phi^\dagger(z^+_k(\lam))\Phi(z^-_k(\lam))\right\ra}{\left\la\prod_{k=0}^{n-1}\Phi^\dagger(z^+_k)\Phi(z^-_k)\right\ra}\rb=\log\lb 1+(C^\Delta_{\phi\phi})^2\sum_{m=1}^{n-1} \frac{(\theta)^{2\Delta}}{\lb c(\theta=0)\sin^2\lb \frac{m\pi}{n}\rb\rb^{\Delta}}+\cdots\rb\nn\\
&&=\lb\frac{\theta}{1-2\lam+\lam^2(1-t^2)}\rb^\Delta\:\sum_{m=1}^{n-1} \frac{1}{\sin^{2\Delta}\lb \frac{\pi m}{n}\rb}+\cdots
\eea
where $\Delta$ is the dimension of the lightest scalar primary. The inverse Laplace transform of the first term with $\lam\to -\lam$ is:
\bea
&&\lb\frac{\theta}{1+2\lam+\lam^2(1-t^2)}\rb^\Delta=\int_0^\infty ds\: e^{-s\lam}\: \mu_t(s)\nn\\
&&\mu_0(s)=\frac{e^{-s} s^{2\Delta-1}}{\Gamma(2\Delta)}\nn\\
&&\mu_1(s)=\frac{e^{-s/2}s^{\Delta-1}}{2^\Delta \Gamma(\Delta)},
\eea
which implies that all derivatives of the sandwiched R{\'e}nyi divergence are non-negative for spatial and null deformations at this order in perturbation theory.

One can go to higher orders in perturbation theory. If we we focus on the case $t=0$ and assume that the second light primary has dimension $\Delta_2>\frac{3}{2}\Delta_1$, at the next order in perturbation theory we have
\bea
\log\lb 1+(C^\Delta_{\phi\phi})^2\sum_{m=1}^{2} \frac{(2\sin\theta)^{2\Delta}}{\lb 4 c(\theta)\sin^2\lb \frac{m\pi}{3}\rb\rb^{\Delta}}+C^\Delta_{\Delta\Delta}\lb \frac{(2\sin\theta)}{3(1-\lam)^2} \rb^{3\Delta}+\cdots \rb
\eea
Since the first order term is non-negative it is hard to use the monotonicity constraint to derive an inequality regarding only the OPE coefficients.\footnote{We thank Shu-Heng Shao for pointing this out to us.}

\section{Discussion and Generalizations}\label{sec:Dis}

In summary, we defined the sandwiched R{\'e}nyi divergence as a R{\'e}nyi generalization of relative entropy in any von Neumann algebra and explored the consequences of its monotonicity for correlation functions of field theory. We found new inequalities for correlation functions and conjectured a constraint on the second in null derivatives of the R{\'e}nyi family that is a generalization of quantum null energy condition. It is interesting to explore the implications of these inequalities in large $N$ and holographic theories. The holographic dual of sandwiched R{\'e}nyi divergences was constructed in \cite{lashkari2016gravitational}. It is natural to ask whether monotonicity can be used to constrain the effective field theory in the bulk. We postpone this to future work.

The physics interpretation of the R{\'e}nyi divergences comes from their connection with the resource theory of thermodynamics. In an out-of-equilibrium quantum system with long-range interactions, there are many independent second laws of thermodynamics that constrain state transformations, each corresponding to a generalized free energy \cite{brandao2015second}. These free energies are precisely the sandwiched R{\'e}nyi divergences and Petz divergences we studied here. This suggests that in conformal field theory and gravity the monotonicity of R{\'e}nyi divergences for $n>1$ can be independent of the monotonicity of relative entropy. In fact, this was shown to be the case in holography in a recent work \cite{bernamonti2018holographic}.

The averaged null energy condition is a universal constraint in all quantum field theories that was recently proved in \cite{faulkner2016modular} using the monotonicity of relative entropy. Later, the authors of \cite{hartman2017averaged} presented an alternative proof that relied on the analyticity and the causality of correlation functions. Our work sheds light on the connection between the two approaches by constructing information-theoretic quantities whose monotonicity is tied to the analyticity of correlation functions.

Finally, we would like to mention a few generalizations of this work. For most of our work here we focused only on the operator monotone function $f(z)=z^\alpha$. An arbitrary operator monotone function can be characterized in the following way:  
A function is operator monotone if and only if it has the representation \cite{schilling2012bernstein}
\bea
f(z)=a z+b +\int_0^\infty d\beta \mu(\beta)\lb \frac{1}{\beta}-\frac{1}{z+\beta}\rb
\eea
for $a,b\geq 0$ and $\mu(\beta)$ a measure that satisfies
\bea
\int_0^1 \frac{d\beta}{\beta} \mu(\beta)+\int_1^\infty \frac{d\beta}{\beta^2}\mu(\beta)<\infty\ .
\eea
For instance, if we take $\mu(\beta)=\frac{\sin(\pi \alpha)}{\pi}\beta^\alpha$ the condition above is achieved if $0<\alpha<1$, and the function is simply $f(z)=z^\alpha$. The measures constructed from arbitrary operator monotone functions in finite quantum systems were studied in \cite{wilde2018optimized}. It would be interesting to see if one can relate the monotonicity of other $f(\Delta_{\Om|\Psi})$ to correlation functions and obtain new constraints.

\section{Acknowledgments}
I am greatly indebted to Edward Witten whose suggestion to consider the non-commutative $L_p$ spaces initiated this work. I would also like to thank Nima Arkani-Hamed, Hong Liu, Raghu Mahajan, Srivatsan Rajagopal, Shu-Heng Shao, Douglas Stanford and Yoh Tanimoto for discussions at various stages of this project. This work was supported by a grant-in-aid (PHY-1606531) from the National Science Foundation.  
\appendix

\section{Proof of Properties Listed for Quasi-entropy}\label{appA0}
In this appendix we prove the the follwoing properties of the Petz quasi-entropy:

\begin{enumerate}
\item If $A\subset B$ and $a\in\mathcal{A}_A$ then it increases monotonically with system size: $$D^A_{\alpha,a}(\Phi\|\Om)\leq D^B_{\alpha,a}(\Phi\|\Om).$$

\item If $U$ is a unitary in $\mathcal{A}_A$ then $D_{\alpha,a}^A(U\Phi\|V\Om)=D_{\alpha,V^\dagger a U}^A(\Phi\|\Om)$. 

\item It increases monotonically in $\alpha$. If $\alpha\leq \beta$ then $D^A_{\alpha,a}(\Phi\|\Om)\leq D^A_{\beta,a}(\Phi\|\Om)$. 

\end{enumerate}

Statements (1) and (2) follows respectively from the monotonicity relation in (\ref{mono}) and the transformation rule (\ref{transformunit}) for the relative modular operator under unitaries in the algebra. 
To show the monotonicity in $\alpha$ we consider the spectral decomposition of the relative modular operator:
\bea
\Delta^{-1}_{\Om|\Phi}= \int_0^\infty \lambda \: P(d\lambda),
\eea
where $P(d\lambda)$ is a projection-valued measure.  
According to Holder's inequality, if $\mu(\lam)$ is a normalized probability measure, $\alpha>0$, $p>1$ and satisfies $\frac{1}{p}+\frac{1}{q}=1$ we have:
\bea
\int \lam^\alpha d\mu(\lam)\leq \lb \int \lam^{p\alpha} d\mu(\lam)\rb^{1/p} \lb \int d\mu(\lam)\rb^{1/q}= \lb \int \lam^{p\alpha} d\mu(\lam)\rb^{1/p}\ .
\eea
Choosing $p=\beta/\alpha>1$ we obtain
\bea
\la\Psi|\Delta_{\Om|\Phi}^{-\alpha}|\Psi\ra&=&
\int \lambda^\alpha \la\Psi|P(d\lambda)|\Psi\ra\leq \lb \int \lambda^{\beta} \la\Psi|P(d\lambda)|\Psi\ra\rb^{\alpha/\beta}=\la\Psi|\Delta_{\Om|\Phi}^{-\beta}|\Psi\ra^{\alpha/\beta}\nn,
\eea
where $|\Psi\ra$ is an arbitrary state. 
This establishes monotonicity in $\alpha$. That is for $0\leq\alpha\leq\beta\leq1$:
\bea
D_{\alpha,a}(\Phi\|\Om)=\frac{1}{\alpha}\log \la\Phi|a^\dagger \Delta_{\Om|\Phi}^{-\alpha}a|\Phi\ra\leq \frac{1}{\beta}\la\Phi|a^\dagger\Delta_{\Om|\Phi}^{-\beta}a|\Phi\ra=D_{\beta,a}(\Phi\|\Om)\ .
\eea
A similar argument using the spectral decomposition of $\Delta_{\Om|\Phi}$ shows that the above equation holds for any $-1\leq \alpha \leq\beta\leq1$. 

\section{Proof of Properties Listed for Petz divergence}\label{appA}

In this appendix, we prove of the properties of the Petz divergence listed below (\ref{Petzdiv}).

\begin{enumerate}

\item It is non-negative and vanishes when $A$ shrinks to zero. 

\item It is invariant under the rotation of both vectors by the same unitary in $A$.

\item At $\alpha=0$ it is smooth and equal to the relative entropy.

\item Under swapping vectors $|\Phi\ra$ and $|\Om\ra$ it satisfies
\bea
D_{-\alpha}(\Phi\|\Om)=\frac{1-\alpha}{\alpha}\: D_{\alpha-1}(\Om\|\Phi)\ .
\eea
\end{enumerate}

Statement (1) follows from the inequality in (\ref{fmono}). Since Petz divergence is monotonic under the restriction to subregions and it vanishes as the region size shrinks to zero, it follows that it is non-negative.
The second statement follows trivially by setting $a=1$ and $U=V$ in property (2) of quasi-entropies.
The limit $\alpha\to 0$ from below and above equal the definition of relative from (\ref{relativeentropy}): 
\bea
\lim_{\alpha\to 0}D_\alpha(\Phi\|\Om)=-\log\la\Phi|\log\Delta_{\Om|\Phi}|\Phi\ra=S(\Phi\|\Om)\ .
\eea

Statement (4) says that under swapping vectors $|\Phi\ra$ and $|\Om\ra$ the Petz divergence satisfies
\bea
D_{-\alpha}(\Phi\|\Om)=\frac{1-\alpha}{\alpha}\: D_{\alpha-1}(\Om\|\Phi)\ .
\eea
Remember that the Petz divergence is invariant under $|\Phi\ra\to U'|\Phi\ra$ with $U'$ a unitary in the $A'$. This freedom can be used intelligently to satisfy $|\Phi\ra=\Delta^{1/2}_{\Phi|\Om}|\Om\ra$. Such a choice is called the vector representative of the state in the ``natural cone". See \cite{Haag:1992hx} for more detail.
Equipped with this fact, the claim follows:
\bea
D_{-\alpha}(\Phi\|\Om)&=&\frac{-1}{\alpha}\log\la\Phi|\Delta_{\Om|\Phi}^{\alpha}|\Phi\ra\nn\\
&=&\frac{1-\alpha}{\alpha}\lb \frac{1}{\alpha-1}\log\la\Om|\Delta^{1/2}_{\Phi|\Om} \Delta_{\Om|\Phi}^{\alpha}\Delta^{1/2}_{\Phi|\Om}|\Om\ra\rb\nn\\
&=&\frac{1-\alpha}{\alpha}\lb \frac{1}{\alpha-1}\log\la\Om|\Delta^{1-\alpha}_{\Phi|\Om}|\Om\ra\rb\nn\\
&=&\frac{1-\alpha}{\alpha}D_{\alpha-1}(\Om\|\Phi)\ .
\eea
At the symmetric point $D_{-1/2}(\Phi\|\Om)=D_{-1/2}(\Om\|\Phi)=-2\log\la\Phi|\Om\ra$. The non-negativity of the Petz divergence implies that $0\leq \la\Phi|\Om\ra\leq 1$ for all normalized vectors in the natural cone. 

\section{Proof of Inequality (\ref{upperlower})}\label{AppB}
The claimed inequality in (\ref{upperlower}) is:
\bea
D_{\frac{\alpha}{1-\alpha}}(\Phi\|\Om)\geq S_\alpha(\Phi\|\Om)\geq D_\alpha(\Phi\|\Om)\ .
\eea

The lower bound follows from the definition of the sandwiched R{\'e}nyi divergences. The upper bound can be written as 
\bea
\la\Phi|\Delta_{\Om|\Phi}^{-\frac{\alpha}{1-\alpha}}|\Om\ra^{\frac{(1-\alpha)}{\alpha}}\geq \sup_\Psi\la \Phi| \Delta_{\Om|\Psi}^{-\alpha}|\Phi\ra^{\frac{1}{\alpha}}\ .
\eea
Using $|\Phi\ra=\Delta_{\Om|\Phi}^{-1/2}|\Om\ra$ it simplifies to
\bea
\|\Delta_{\Om|\Phi}^{-\frac{1}{2(1-\alpha)}}|\Phi\ra\|^{(1-\alpha)}\geq \sup_\Psi \| \Delta_{\Om|\Psi}^{-\frac{\alpha}{2}}|\Phi\ra\|\ .
\eea
Then, we use $(\Delta_{\Om|\Psi}^A)^{-1}=\Delta_{\Phi|\Om}^{A'}$ and $\alpha=1-\frac{2}{p}$ to we write it as
\bea
\|(\Delta^{A'}_{\Phi|\Om})^{\frac{p}{4}}|\Om\ra\|^{\frac{2}{p}}\geq \sup_\Psi \|(\Delta^{A'}_{\Psi|\Om})^{\frac{1}{2}-\frac{2}{p}}|\Phi\ra\|\ .
\eea
This inequality is a generalization of the Araki-Lieb-Thirring inequality \cite{lieb1991inequalities} to von Neumann algebras that can be proved using the interpolation theory of non-commutative $L_p$ spaces. However, this goes beyond the scope of this work and we refer the interested reader to the proof presented in Theorem 12 of \cite{berta2018renyi}.

\bibliographystyle{JHEP}

\bibliography{monotone}
\end{document}